\documentstyle[12pt,cite,epsfig]{article}

\renewcommand{\today}{18 July 1997 \\
Revised: 4 December 1997}

\newcommand{\nc}{\newcommand}
\nc{\be}{\begin{equation}}
\nc{\ee}{\end{equation}}
\nc{\bea}{\begin{eqnarray}}
\nc{\eea}{\end{eqnarray}}
\nc{\beas}{\begin{eqnarray*}}
\nc{\eeas}{\end{eqnarray*}}
\nc{\noi}{\noindent}
\nc{\sD}{\not \! \! D}
\nc{\s}[1]{\not \! #1}
\nc{\non}{\nonumber}
\nc{\bb}{\bibitem}
\nc{\lf}{\left}
\nc{\ri}{\right}
\nc{\mb}[1]{\makebox[#1]{}}
\nc{\pa}{\partial}
\nc{\sA}{\not \! \! A}
\nc{\newsec}[1]{\section{#1}\mb{0.5cm}}
\nc{\h}{\frac{1}{2}}
\nc{\ra}{\rightarrow}
\nc{\la}{\leftarrow}
\nc{\ep}{$e^+e^-\ra\pi^+\pi^-\;$}
\nc{\emuon}{$e^+e^-\ra\mu^+\mu^-\;$}
\nc{\epp}{$e^+e^-\ra\pi^+\pi^0\pi^-\;$}
\nc{\elec}{$e^+e^-\ra\gamma^*\ra e^+e^-\;$}
\def\mathunderaccent#1{\let\theaccent#1\mathpalette\putaccentunder}
\def\putaccentunder#1#2{\oalign{$#1#2$\crcr\hidewidth
\vbox to.2ex{\hbox{$#1\theaccent{}$}\vss}\hidewidth}}

\nc{\ti}{\mathunderaccent\tilde}
\nc{\M}{{\cal M}}
\nc{\rw}{$\rho\!-\!\omega\;$}

\begin{document}
\def\hhha{\rule[-3.mm]{0.mm}{7.mm}}
\def\hhhb{\rule[-3.mm]{0.mm}{12.mm}}
\thispagestyle{empty}
\begin{flushright}
ADP-97-14/T251 \\
UK-97-16 \\
LPNHE 97-02\\
hep-ph/9707509

\end{flushright}

\begin{center}
{\large{\bf New Results in $\rho^0$ Meson Physics}}\\
{\bf [Eur. Phys. J. C 2 (1998) 269-286]}\\
\vspace{2.2 cm}M.~Benayoun$^{a}$, S. Eidelman$^{b}$, K.~Maltman$^{c,d,e}$,
 H.B. O'Connell$^{d,f}$, B.~Shwartz$^{b}$ and A.G.~Williams$^{d,e}$\\
\vspace{1.2 cm}
{\it $^{a}$ LPNHE des Universit\'es Paris VI et VII--IN2P3, Paris,
France}\\
\vspace{.5 cm}
$^{b}$ {\it Budker Institute of Nuclear Physics, Novosibirsk 630090, Russia} \\
\vspace{.5cm}
$^{c}${\it Mathematics and Statistics, York University, 4700 Keele St., \\
North York, Ontario, Canada M3J 1P3 }\footnote{Permanent address.}\\
\vspace{.5 cm}
$^{d}${\it Department of Physics and Mathematical
Physics \\ University of Adelaide 5005, Australia } \\
\vspace{.5 cm}
{\it
$^{e}$Special Research Centre for the Subatomic Structure of Matter, \\
University of Adelaide 5005,
Australia } \\
\vspace{.5 cm}
$^{f}${\it Department of Physics and Astronomy,
University of Kentucky\\
Lexington, KY 40506, USA }\footnote{Present address.} \\
\vspace{1.2 cm} \today
\vspace{1.2 cm}
\begin{abstract}
We compare the predictions of a range of existing models based
on the Vector Meson Dominance hypothesis with data on
$e^+e^-\rightarrow \pi^+\pi^-$ and $e^+e^-\rightarrow \mu^+\mu^-$
cross-sections and the phase and near-threshold behavior of the timelike
pion form factor, with the aim of determining which (if any) of these
models is capable of providing an accurate representation of the full
range of experimental data.  We find that, of the models considered,
only that proposed by Bando {\it et al.}
is able to  consistently account for all
information, provided one allows its  parameter $a$ to vary from
the usual value of 2 to 2.4.
Our fit with this model gives
a point--like coupling $\gamma \pi^+ \pi^-$
of magnitude $\simeq -e/6$, while the common formulation of
VMD excludes such a term. The resulting values
for the $\rho$ mass and $\pi^+\pi^-$ and $e^+e^-$
partial widths as well as the branching ratio for the decay
$\omega \rightarrow \pi^+\pi^-$ obtained within the context of this model are
consistent with previous results.

\end{abstract}
\end{center}
\vspace{2.5cm}
\begin{flushleft}
E-mail: {\it benayoun@in2p3.fr; eidelman@vxcern.cern.ch;
maltman@fewbody.phys.yorku.ca
hoconnel@pa.uky.edu;
awilliam@physics.adelaide.edu.au
 }
\\ Keywords: vector mesons, photon, muons.\\
PACS: 11.80.Et Partial-wave analysis, 12.40.Vv Vector-meson
dominance.

\vfill
{\it }
\end{flushleft}

\newpage

\pagenumbering{arabic}
\section{Introduction.}

Our aim is to study  the various ways to describe the $\rho^0$
meson in order to find an optimum modelling able to account
 most precisely  for  the
known features of the physics involving this meson.
This is mainly motivated by the fact that a precise knowledge
of its properties is of fundamental importance in several
fields of particle physics.
It is important to emphasise that our philosophy is to look for
the simplest models as these are the most useful in
application to other systems, due to their ease of implementation.
Naturally, such models should, as much as possible, respect
basic general principles such as gauge invariance and unitarity.
One must keep in mind that any parameters quoted for a given
model are relevant only to that model.
Indeed, a study of the model-dependence of resonance parameters is one
of the principal goals of this work.  It should also be noted that while
each of these models is related to some underlying effective field
theory through an effective Lagrangian, the models we are using here are
simple amplitudes arising from an assumption of almost
complete $s$-channel resonance
saturation.  The appropriateness of this assumption away from the
resonance region can only be judged by quantitative studies of higher
order (e.g., loop) effects in the corresponding effective field theories.
While this is clearly a very important task, it is not our concern here
and will not be considered further.

For this purpose, we study the strong
interaction corrections to one-photon mediated processes
in the low energy region where QCD is non-perturbative.
To do this we shall look at two related processes, \ep and \emuon.
The effect of the
strong interaction is obvious in the first reaction and provides a
large enhancement to the
production of pions in the vector meson resonance region
\cite{orsay,benaksas,barkov,kurdadze}.
This enhancement,
relative to what would be expected for a structureless, pointlike
pion, is reflected in the deviation of the pion form factor, $F_\pi (q^2)$,
from $1$, and is primarily associated with the $\rho$ meson (where
$q_\mu$ is the four momentum of the virtual photon).
This form factor is successfully modelled in the intermediate energy
region using the vector meson dominance (VMD) model
\cite{review}. VMD assumes that the photon interacts with
physical hadrons through vector mesons and it is these
mesons that give rise to the enhancement, through their resonant
(possessing a complex pole) propagators of the form
\be
D_{\mu\nu}(q^2)=\frac{-g_{\mu\nu}}{q^2-{m}_V^2+i{m}_V\Gamma_V
(q^2)},
\ee
where ${m}_V$ and $\Gamma_V$ are the (real valued) mass and
the momentum-dependent width.
(Here we have included only that part of the propagator
which survives when coupled to conserved currents.)

Traditionally, VMD assumes that {\em all} photon--hadron coupling
is mediated by vector mesons.
However, from an empirical point of view,
one has the freedom, motivated by Chiral Perturbation Theory (ChPT)
to include other contributions to such interactions. For instance, in a
fit to $F_\pi(q^2)$ \cite{benayoun}, it was shown that
a non-resonant photon--hadron coupling can be accommodated
merely by shifting the
mass and width of the $\rho^0$ by about 10 MeV.  Thus, the values
extracted from $F_\pi(q^2)$ for the
$\rho$ mass, $m_\rho$, and width, $\Gamma_\rho$, are
model-dependent and in quoting values
for them, the model used should be clearly stated.  We note
in passing that to define vector meson masses and widths in a
process-independent way, one should refer to the location of the corresponding
complex pole in the S-matrix.  One should, however,
bear in mind that alternate, and more traditional definitions of the mass
and width not tied to the location of the S-matrix pole
(for example a definition of $m_\rho^2$ as that value of $q^2$ for
which the $P$--wave $\pi\pi$ phase shift passes through $90^\circ$)
are in general specific to the process employed in the definition.
The process-dependence of such
alternate mass and width definitions has in
fact led a number of authors to advocate using the S-matrix pole
position to provide a process-independent definition of the $Z^0$
mass and width in the Standard Model\cite{poles2}.

Naturally, VMD can be applied to many other systems. We can consider the
process $\eta' / \eta \ra \pi^+\pi^-\gamma$ taking place through
a combination
of resonant (such as $\eta' / \eta \ra \rho\gamma\ra \pi^+\pi^-\gamma$)
and non-resonant channels. In this manner an acceptable fit to
data can be achieved with a range of combinations for the $\rho$ parameters
and non-resonant terms \cite{benayoun,benayoun2}. Recent interest in this process
has centred on the non-resonant term, which, if it arises from anomalous
box and triangle  diagrams,
provides a possible test of QCD \cite{chan1,chan2,WZ}. However, to
determine the size of any non-resonant contribution, the
resonant meson parameters need to be well
fixed \cite{benayoun,benayoun2}, and  thus $F_\pi(q^2)$  well understood.
This is one of the main aims of this paper.

We thus turn our attention to \emuon. In modelling the strong
interaction correction to the photon propagator, VMD assumes that the
strong interaction contribution is saturated by the
spectrum of vector meson resonances \cite{W}.  Therefore,
in principle, we can extract information on the vector meson parameters
(independently of the \ep fit) without having to worry
about non-resonant processes. However, as the vector mesons enter
in \emuon with an extra
factor of $\alpha$ compared with \ep, their contributions are considerably
suppressed,
making their extraction difficult.  For this reason, we shall perform a {\em
simultaneous} fit to both sets of data, in order to impose the best possible
constraint on the vector meson parameters, and see if existing muon data are
already precise enough in order to constrain the $\rho^0$ parametrisation.

Another way to constrain the  descriptions of the
$\rho^0$ meson is to compare the strong interaction
$\pi \pi$ phase obtained using the various VMD parametrisations determined
in fitting \ep with the corresponding phase \cite{petersen} obtained
using  $\pi \pi$ scattering
{data} and the general principles
of quantum field theory, as well as the near--threshold predictions of
ChPT. This happens to be more fruitful and conclusive
in showing how VMD should be dealt with in order to reach an agreement
with a large set of data and with the basic principles of quantum field
theory.

The hadronic dressing of the photon propagator (the one-photon irreducible
self-energy $\Pi^{\rm had}(q^2)$) is also of interest for another reason. The
anomalous magnetic moment of the muon can now be measured to such
accuracy \cite{muong}
that the strong interaction correction is important \cite{BW}.  This
needs to be completely understood if one is to look for physics beyond the
Standard Model in
this quantity. At present the correction is inferred from
$\sigma(e^+e^-\ra$ hadrons) using dispersion theory and the optical theorem
or estimated with hadronic models (which, being
non-perturbative, are difficult to use). The
process \emuon allows for a {\em direct} examination of the
strong interaction
modification  to the photon propagator. Ideally we would
obtain the low energy
corrections to the photon propagator
experimentally, and be able to use QCD perturbatively
for higher energies  (though the threshold above which we can ignore
non-perturbative effects
is difficult to determine \cite{roberts}).
The $\phi$ meson has already been seen
in \emuon \cite{orsay,kurdadze}. It is only
noticeable around the pole region where, due to its small width (4.4 MeV), it
produces a  sharp peak
easily seen in the available data. The large width of the
$\rho$ further suppresses the $\rho$ and $\omega$ contributions.

The outline of the present work is as follows. In section \ref{models}
we describe the various formulations of the VMD assumption and
the unitarisation procedure; we also discuss  the
phase definition  relevant for our purpose.   The fit procedure is
sketched in section \ref{fitcond} and  implemented
in section \ref{fitee}.
Comparison with the isospin 1 $P$--wave $\pi \pi$ phase shift deduced from
$\pi \pi$ scattering theory is the subject of section \ref{phaseshift},
while in section \ref{threshold} we give the near--threshold  parameter
values that are deduced from our fits of the \ep cross section and they are
compared to predictions and to other experimental determinations. The
results obtained are discussed in  section \ref{discussion} where we
present our optimal fit for the $\rho^0$ parameters  and
the model which accounts for its properties the most appropriately.
Finally, we summarise our conclusions in section \ref{conclusion}.

\section{Vector meson models}\label{models}

We shall now provide a description of the various
models we will use to fit
the data for both \ep and \emuon. The cross-section for \ep
is given by (neglecting the electron mass)
\be
\sigma=\frac{\pi \alpha^2}{3}\frac{(q^2-4m_\pi^2)^{3/2}}{(q^2)^{5/2}}
|F_\pi(q^2)|^2,
\label{tempex}
\ee
where the form factor, $F_\pi(q^2)$ is determined by the specific model.
Similarly, $F_\mu(q^2)$ is defined to be the form factor
for the muon, and the cross-section for \emuon is given by,
\be
\sigma=
\frac{4\pi\alpha^2}{3q^2}\sqrt{1-\frac{4m_\mu^2}{q^2}}
\left(1+\frac{2m_\mu^2}{q^2}\right)|F_\mu(q^2)|^2.
\label{cross}
\ee
It is worth noting that these standard definitions of $F_\pi(q^2)$ and
$F_\mu(q^2)$ contain all non-perturbative effects, including for example
the photon vacuum polarisation, since Eqs.~(\ref{tempex}) and (\ref{cross})
are written assuming a perturbative photon propagator.

We shall use VMD, the Hidden Local Symmetry model of Ref. \cite{bando}
(hereafter refered to as HLS) and
what we will refer to as the WCCZW model (a phenomenological modification
of the general framework of Ref.~\cite{WCCWZ}, based on work
by Birse\cite{birse}), as well as modifications of these models,
which cover or underlie a large class of effective Lagrangians
describing the interactions of photons, leptons and pseudoscalar and vector
mesons. Birse \cite{birse} has shown that typical effective theories
involving vector mesons based on a Lagrangian approach, such as
``massive Yang-Mills" and ``hidden gauge" (i.e., HLS-type) are equivalent.
We therefore
consider our following examination to be reasonably comprehensive.
Numerical approaches to the strong interaction in the vector meson
energy region \cite{num}, which are not based
on an effective Lagrangian involving meson degrees of freedom
do not yet possess the required calculational accuracy for our task.

\subsection{VMD}
The simplest model is VMD itself. As has been discussed
in detail elsewhere
\cite{review,OWBK,OPTW2} VMD has two equivalent formulations, which we
shall call VMD1 and VMD2.
The VMD1 model has a momentum-dependent coupling between the photon and the
vector mesons and a direct coupling of the photon to the hadronic final
state. The resulting form factor is (to leading order in isospin
violation and $\alpha = e^2/4\pi$):
\bea\non
F_\pi^{\rm VMD1}(q^2)&=&1-g^{\rm VMD1}_{\rho\gamma}(q^2)
\frac{g_{\rho\pi\pi}}{[q^2-m^2_\rho+im_\rho
\Gamma_\rho(q^2)]}\\
&&-g^{\rm VMD1}_{\omega\gamma}(q^2)
\frac{1}{[q^2-m^2_\omega+im_\omega
\Gamma_\omega(q^2)]}Ae^{i\phi_1}.
\label{ff1}
\eea

The $\omega$ enters into the isospin 1 \ep interaction with an attenuation
factor specified by
the pure real $A$ and the Orsay phase, $\phi$ \cite{MOW}; these
can be extracted from experiment\footnote{Note in Ref.~\cite{MOW}
that $A$ and $\phi$ were defined through the S matrix pole positions
(equivalent to Eq.~(\ref{ff2}) with {\em constant} widths).
In our fit procedure, $A$  is connected with the
width $\Gamma(\omega \rightarrow \pi^+ \pi^-)$ (see Ref.\ \cite{benayoun}).}
For VMD2 we have,
\be
F_\pi^{\rm VMD2}(q^2)=-g^{\rm VMD2}_{\rho\gamma}
\frac{g_{\rho\pi\pi}}{[q^2-m^2_\rho+im_\rho
\Gamma_\rho(q^2)]}-g^{\rm VMD2}_{\omega\gamma}\frac{1}
{[q^2-m^2_\omega+im_\omega
\Gamma_\omega(q^2)]}Ae^{i\phi_2}.
\label{ff2}
\ee
The form factor for the muon, however, in both representations
(i=1, 2) is given by
\be
F_\mu^{\rm VMDi}=1+\sum_V e^2 [g_{V\gamma}^{\rm VMDi}(q^2)]^2
\frac{1}{q^2-{m}_V^2+i{m}_V\Gamma_V(q^2)}\frac{1}{q^2} ,
\label{muonff}
\ee
which is consistent with previous expressions  for the
$\phi$-meson \cite{orsay,PR}.
In higher order ({\it i.e.}, in all but the minimal VMD picture)
there will also be contributions from non--resonant processes (such
as two--pion loops), but these are expected to be small near resonance and
the non--resonant background is, in any case, fitted in extractions of
resonance parameters from the experimental data.
The photon--meson coupling,
$eg_{V\gamma}$ is fixed in VMD2 by \cite{PR} \be \Gamma_{V\ra
e^+e^-}=\frac{4\pi\alpha^2}{3{m}_V^3}g^2_{V\gamma}.
\label{coupdef}\ee
The (dimensionless) universality coupling, $g_V$, is then defined by
\be
g_{V\gamma}^{\rm VMD2}={m}_V^2/g_V\label{vmd2g}
\ee for VMD2 \cite{OWBK,MOW}.
This coupling (and universality) has been most closely studied
for the $\rho$ meson. A gauge-like argument \cite{review,Sak}
suggests that the $\rho$ couples to all
hadrons with the same strength  $g_V$ (universality)
\cite{OWBK}. However, experimentally, universality is observed to be not
quite exact \cite{klingl},
so we introduce the quantity $\epsilon$ (to be fitted)
through
\be
g_{\rho\gamma}^{\rm VMD2}=\frac{{m}_{\rho}^2}{g_{\rho\pi \pi}} (1 +\epsilon)
\label{epsy}
\ee
where $g_{\rho\pi \pi}$ and $\epsilon$ are to be extracted from
the fit to \ep.
For VMD1, it can be seen that the photon-meson coupling results from
replacing the mass term in Eq.~(\ref{vmd2g}) by $q^2$\cite{review,Sak}.
In this case Eq (\ref{epsy}) should be replaced by
\be
g_{V\gamma}^{\rm VMD1}=\frac{q^2}{g_{\rho\pi \pi}} (1 +\epsilon)
\label{1vgam}~~.
\ee

One can
easily see that in VMD1 the hadronic correction to the photon propagator
goes like $q^4$ and so maintains the photon pole at $q^2=0$. For VMD2,
it is not obvious \cite{Sak_orig} that gauge invariance
is maintained until one considers the inclusion of
a bare photon mass term in the VMD Lagrangian that
exactly cancels the hadronic correction at $q^2=0$.
This argument \cite{KLZ67} assumes a $\rho-\gamma$
coupling of the form $em_\rho^2/g_{\rho}$ and a
bare photon mass ($e^2m_\rho^2
/g_{\rho}^2$) (the calculation is presented in detail in Ref.~\cite{review}).
In the presence of a finite $\epsilon$, as in Eq.~(\ref{epsy}),
gauge invariance is similarly preserved by including a photon mass
term ($e^2m_\rho^2(1+\epsilon)^2/g_{\rho}^2$) leading to a massless
photon as expected \cite{craig}.

The presence of a finite $\epsilon$
does affect the charge normalisation condition
$F_\pi(0)=1$, but this is merely an artifact of
the simple $\rho$ propagators we are considering. A
more sophisticated  version, such as used by Gounaris and Sakurai \cite{GS}
which fully accounts for below threshold behaviour,
maintains $F_\pi(0)=1$ in the presence of $\epsilon\neq0$.
One could, alternatively, include an $s$ dependence to $\epsilon$ such
that $\epsilon(0)=0$.
In any case, the phenomenological significance for
the physical $\rho$ (for the data region we are fitting) is
negligible, we achieve an excellent fit to the data with the simple
form we use but are not advocating its use outside this fitting region,
namely above the two-pion threshold.

The choice of the detailed form of the momentum-dependent
width, $\Gamma_V(q^2)$, allows one certain amount of
freedom. As the complex poles of the amplitude are field-choice
and process-independent properties of the S-matrix  \cite{poles1}
one could, on the one hand, expand the propagator as a Laurent
series in which the non-pole terms go into the background
\cite{bernicha}. Alternatively, one could use the $l-$wave
momentum-dependent width to account for the branch point structure of
the propagator above threshold ($q^2=4m_\pi^2$) \cite{benayoun,PRoos,GS}.
This form for the momentum dependent width arises naturally from the
dressing of the $\rho$ propagator in an appropriate
Lagrangian based model (see Sec.\ 5.1 of
Klingl {\it et al.} for a detailed treatment \cite{klingl}) and, for such
models, is given by
\be
\Gamma_\rho(q^2)=\Gamma_\rho\left[\frac{p_{\pi}(q^2)}{p_{\pi}(m_{\rho}^2)}
\right]^3\left[\frac{m_\rho^2}{q^2}\right]^{\lambda/2},
\label{mesnwid}
\ee
introducing the fitting parameters $\Gamma_{\rho}$
(the width of the $\rho^0$ meson at $q^2=m_{\rho}^2$) and $\lambda$, which
generalises the usual $l$--wave expression \cite{benayoun}
to model the fall--off of the $\rho$ mass distribution;
the usual case ({\it i.e.} $\lambda=1$) is associated
with a $\rho$ coupling to pions of the form
$g_{\rho\pi\pi}\rho^{\mu} (\pi^+ \partial_{\mu}
\pi^- - \pi^- \partial_{\mu} \pi^+)$, with $g_{\rho\pi\pi}$
independent of $q^2$,
as shown in Ref. \cite{klingl}. Note that the parameters $\lambda$,
$m_{\rho}$ and $\Gamma_{\rho}$ are model-dependent (as we see in the tables of
results, different models yield different values for these parameters).
Note that in Eq.~(\ref{mesnwid})
we have defined the pion momentum in the centre of mass system
\be
p_{\pi}(q^2)=\frac{1}{2} \sqrt{q^2-4m_{\pi}^2}.
\label{ppi}
\ee

Before closing this section, let us remark that the pion form
factor associated with VMD1 (Eq. (\ref{ff1})) fulfills
automatically the condition
$F_{\pi}(0)=1$ whatever the value of the universality violating
parameter $\epsilon$ (see Eq. (\ref{1vgam})). This is not the case for
the pion form factor associated with VMD2 (Eq. (\ref{ff2})), as  can
be seen from Eqs. (\ref{ff2}) and (\ref{epsy}). Here,  we will concern
ourselves exclusively with fitting data in the above threshold region
and simply note it is a relatively straightforward matter to generalise
the VMD models considered to satisfy this condition.  Detailed
considerations of this issue are left for future work.

While of course VMD1 and VMD2 are equivalent in the limit of
exact universality if one keeps {\it all} diagrams, ({\it i.e.},
if one works to infinite order in perturbation theory), in
any practical calculation one cannot do that and so, in practice,
these two expressions of VMD can give different predictions in
general, even if exact universality is imposed. Moreover, if one releases
(as we do) this last constraint, equivalence of VMD1 and VMD2
is not guaranted, even in principle.

\subsection{The HLS Model}
The Hidden Local Symmetry
(HLS) model \cite{review,bando} introduces a parameter $a$
for the $\rho$ meson within a dynamical symmetry breaking
model framework. This $a$ relates the
constant $g_{\rho\pi\pi}$ to the universality coupling
$g_\rho$ via
\be
g_{\rho\pi\pi}=\frac{a g_\rho}{2}.
\ee
The resulting form factor for the pion is
\be
F_\pi(q^2)=-\frac{a}{2} + 1
- g_{\rho\gamma}\frac{g_{\rho\pi\pi}}
{(q^2-m_\rho^2+im_\rho\Gamma_\rho(q^2))} - g_{\omega\gamma}\frac{1}
{q^2-m^2_\omega+im_\omega
\Gamma_\omega(q^2)}Ae^{i\phi}.\label{fbando} \\
\ee
The original HLS model preserved isospin symmetry and so did not
include the $\omega$. Isospin breaking has recently been studied in a
generalisation of the
HLS model \cite{hash}, however, here we have for simplicity employed
the same $\omega$ terms as  used for VMD.
The relations equivalent to Eqs.~(\ref{vmd2g}) and (\ref{epsy})
for the $\rho$ meson are now
\be
g_{\rho \gamma}=\frac{a}{2} \frac{m_{\rho}^2}{g_{\rho \pi
\pi}}=\frac{m_{\rho}^2}{g_\rho}\ .
\label{epsa}
\ee
We see that setting $a=2$ reproduces VMD2 in the limit of exact universality.
However, we wish to keep $a$ as a free parameter which we can fit to the data.
Note that in the HLS model universality violation and the existence of
a non--resonant coupling $\gamma \pi^+ \pi^-$ are related.
Note also that universality violation can be introduced
in the HLS model without violating the constraint
$F_\pi (0)=1$, in a natural way.

The muon form factor for the HLS model is exactly the same as for VMD2
(see Eqs.~(\ref{muonff}) and (\ref{epsa})).

\subsection{WCCWZ Lagrangian}
Birse has recently discussed \cite{birse} the pion form factor arising from
the WCCWZ Lagrangian \cite{WCCWZ} in which the vector and
axial vector fields transform homogeneously under non-linear chiral
symmetry. The scheme imposes no constraints on the couplings of the
spin 1 particles beyond those of approximate chiral symmetry. Birse's version
of the form factor is (isospin violation is not considered)
\be
F_\pi(q^2)=1-\frac{g_1f_1}{f_\pi^2}\frac{q^4}{q^2-m^2_\rho+im_\rho
\Gamma_\rho(q^2)}+\frac{f_2}{f_\pi^2}q^2,
\label{wccwzo}
\ee
where the first two terms on the RHS are those arising from the
WCCWZ Lagrangian.
The $q^4$ piece grows at large $q^2$ in a way incompatible with
QCD predictions (for a discussion of matching the asymptotic prediction
to a low energy model see Geshkenbein \cite{Gesh}).
 The $f_2$ contribution has been added by Birse to modify
this high energy behaviour toward that expected in QCD. To this end,
Birse sets
\be
f_2=g_1f_1=\frac{f_\pi^2}{m^2_\rho}\label{birsec}
\ee
and recovers the universality limit of the form factor
in which VMD1 and VMD2 are equivalent
(in the zero width approximation).  Note that a $q^2$-dependence of
the non-resonant background is what one would, in general,
obtain from the WCCWZ framework, implemented in its most general form,
which relies only on the symmetries
of QCD.  In constructing a phenomenological implementation we have, however,
simplified the most general form, adding what amounts to
a minimal $q^2$-dependence to the background term of VMD1.  The
resulting form factor, which we will refer to as the WCCWZ model, is then
\be
F_\pi(q^2)=1+b q^2-\frac{g_{\rho\gamma}^{\rm WCCWZ}
(q^2)g_{\rho\pi\pi}}{q^2-m^2_\rho+im_\rho
\Gamma_\rho(q^2)} - g_{\omega \gamma}
\frac{Ae^{i\phi}}{q^2-m_\omega^2+im_\omega\Gamma_\omega(q^2)},
\label{WZM1}
\ee
where we keep $b$ as an independent parameter to be fit.

The WCCWZ model, thus, has one more free parameter
than VMD1. We have added
the $\omega$ contribution as above for all other models. The muon
form factor is exactly the same as for VMD1 (see Eq.~(\ref{muonff})).

It is important to note that the WCCWZ model allows one
to have a non--resonant
term which can be mass dependent, cf., VMD2 which
carries only resonant contributions or VMD1 or the HLS models
which both exhibit only constant non--resonant contributions to the
pion form factor. The expression in Eq.~(\ref{WZM1}) exhibits an unphysical
high energy behaviour; however, as we are only interested in the
pion form factor at low energies (below 1 GeV), this feature is not relevant.
Of course, one can consider that such a polynomial structure at low
energies represents an approximation in the resonance region to
a function going to zero at high energies
{\footnote {For instance, $1+bq^2$ can be considered as
the first terms of the Taylor expansion of a function like $1/(1-bq^2)$
as suggested by \cite{chung}.}}.

\indent The model VMD2 contains only resonant contributions, whereas VMD1,
HLS and WCCWZ also contain a non--resonant part. For VMD1 and HLS, this
term is constant ({\it i.e.}, pointlike) as in standard lowest order
QED. We shall frequently refer to this term as a direct $\gamma\pi\pi$
contribution or coupling.  In the case
of WCCWZ, this non--resonant term also contains a $q^2$--dependent piece
which clearly indicates a departure from a point--like coupling; nevertheless,
we shall also  refer to it as
a direct $\gamma\pi\pi$ coupling for convenience.

\subsection{Elastic Unitarity}
\label{unit}

{ When $\lambda\neq1$}
the pion form factor described above in the VMD, HLS  and WCCWZ
models does not in general {\em exactly} fulfill unitarity.
This is because  one employs simultaneously
a bare, undressed $\rho^0\pi^+\pi^-$ coupling (given by the tree--level
coupling, $g_{\rho\pi\pi}$) and a dressed $\rho^0$ propagator, as
signalled by the presence of a momentum--dependent width in Eq.~(\ref{mesnwid})
(see, {\it e.g.}, Ref.~\cite{klingl}).  To see why this generally creates a
problem with unitarity, consider the $ \pi \pi \ra  \pi \pi$
$P$--wave $I=1$ strong interaction amplitude. It is most convenient
here to employ the $N/D$ formalism for the amplitude T
(see, {\it e.g.}, Ref.~\cite{chew}).

It has been known for some time that the $I=l=1$ $\pi \pi$ amplitude
is consistent with being purely elastic from threshold up
to $\simeq  1$
GeV \cite{estabrooks,baton,proto}. Therefore, in this region, we can
 write{\footnote
{The notation here is $\delta^I_l$.}}

\begin{equation}
S=1+2ip_{\pi}T=\exp{[2i\delta^1_1]}
\label{unit1}
\end{equation}

\noindent from which it follows that

\begin{equation}
T=\displaystyle \frac{\exp{[i\delta^1_1]} ~\sin\delta^1_1}{p_{\pi}}=
\displaystyle \frac{N}{D}
\label{unit2}
\end{equation}

\noindent where $p_{\pi}$ has been defined in Eq.~(\ref{ppi}).
In the region where the $\rho^0$ meson essentially saturates the
$I=l=1$ $\pi \pi$  wave,  the $D$ function may be
approximated by the inverse $\rho^0$ propagator
($s \equiv q^2$),
\be
{D}(s)=s-m_{\rho}^2+im_{\rho} \Gamma_{\rho}(s) ~~~.
\label{unit3}
\ee

$D$ in Eq. (\ref{unit2}), is an analytic function of $s$, having as sole
singularity
in the physical sheet a cut along the real axis ($s \geq 4 m_{\pi}^2$).
Correspondingly, the singularities of the
$N$ function on the physical sheet are all located on the real axis
at {\em negative} values of  $s$ (for example,
singularities produced by exchanges in the $t$ and $u$ channels
projected out onto the $P$--wave). Moreover $N$ is real on the real
axis above threshold ($4 m_{\pi}^2$).
It should also be noted that the $N$ and $D$ functions in Rel.
(\ref{unit2}), are defined up to a multiplicative arbitrary function
$f(s)$, meromorphic in the complex $s$--plane. The choice  used in
 Eq.~(\ref{unit3}) corresponds to a particular
choice of $f(s)$.  Put simply, the
content of Eq.~(\ref{unit3}) is that the phase of $D$ is well
approximated by the phase of the $\rho^0$ propagator in the vector meson
resonance region.

As a consequence of unitarity and
analyticity, $N$ is connected with
the  discontinuity of $D$  across the physical region ($s \geq 4 m_{\pi}^2$)
through
\be
\lim_{\varepsilon \rightarrow 0}~~ [{D}(s-i\varepsilon)-
{D}(s+i\varepsilon)]=2 i  p_{\pi} {N}(s),\ee
\noindent which gives \cite{benayoun},
\be
N(s)=-\frac{m_{\rho} \Gamma_{\rho}(s)}{p_{\pi}}.
\label{unit4}
\ee
As the $N$ function is closely
connected with the $\rho \pi^+ \pi^-$ vertex,
this last relation implies some dressing of the vertex coupling. This
relation illustrates the effect of the parameter $\lambda$,
whose role is simply to model the
contributions from the left--hand singularities
of the scattering amplitude.  Then using Eqs.~(\ref{unit3} -- \ref{unit4}),
it is easy to check that $|S|=1$ is automatically satisfied
in the physical region below $\simeq 1$  GeV, as required by unitarity, and that

\begin{equation}
\tan{\delta^1_1}=\displaystyle \frac{m_{\rho} \Gamma_{\rho}(s)}
{m_{\rho}^2-s}~~.
\label{unit5}
\end{equation}
\noindent
Thus from Eq.~(\ref{unit5}) we can conclude that $\delta^1_1$ can be
well approximated by the negative of the $\rho$ propagator phase,
if one considers resonant contributions only.

One can then show\cite{benayoun} that unitarity can be restored
to the model treatments above by replacing
the coupling $g_{\rho \pi \pi}$ in Eqs.~(\ref{ff1}), (\ref{ff2}),
(\ref{fbando})
and (\ref{WZM1}) with
\be
G_{\rho}(q^2)=\displaystyle
\sqrt{6 \pi
 \frac{m_{\rho}q}{p_{\pi}^3(q^2)}
\Gamma_{ \rho \ra \pi^+ \pi^-}(q^2)}
\label{unit6}
\ee
where $q\equiv\sqrt{q^2}$.  This replacement leads to the
unitarised versions of our VMD models.
The connection between this ``dressed"
vertex function and $g_{\rho \pi \pi}$ gives (cf., Eq.~(4.16) of
Ref.~\cite{klingl})
\be
g_{\rho \pi \pi}=\sqrt{6 \pi
\frac{ m_{\rho}^2}{p_{\pi}^3(m_{\rho}^2)}\Gamma_{\rho}} ,
\label{vtx1}
\ee
from which we see that
\be
g_{\rho \pi \pi}=   G_{\rho}(m_{\rho}^2).
\label{vtx2}
\ee

It should be noted that the left hand side of Eq.~(\ref{unit6})
becomes constant -- and then coincides with $ g_{\rho \pi \pi}$ --
if and only if
$\lambda=1$ (see Eq.\ (\ref{mesnwid})). Therefore, if  $\lambda \equiv 1$
Eqs.~(\ref{ff1}), (\ref{ff2}), (\ref{fbando}) and (\ref{WZM1})
are already unitarised and the singularities
of $N$ (see Eq.~(\ref{unit4})) are simply a branch point going from $s=0$
to $-\infty$, where $s=0$ is also a pole.

Strictly speaking, one should similarly unitarise
the $\omega$ contribution.  However, since the $\omega$ is very narrow,
its contributions are significant only over a very limited range of
$q^2$.  Hence, it is sufficient to employ a constant $\omega\pi\pi$
coupling (or equivalently, constant $A$) in Eqs.~(\ref{ff1}), (\ref{ff2}),
(\ref{fbando}) and (\ref{WZM1}).
The value of $A$ may be determined from the
branching fraction of $\omega\rightarrow\pi^+\pi^-$, as was done in
Ref.~\cite{benayoun} or fitted.
Therefore, the only actual free parameter in the $\omega$ contribution
is the Orsay phase, $\phi$.
This will in no way affect the description of the $\rho^0$ itself, which
is our main concern. Finally,
unitarisation clearly does not affect the expressions for the muon form factor.

The unitarised version of VMD2  coincides with the phenomenological
model called M$_2$ in Refs.~\cite{benayoun,benayoun2}. Here, however,
we shall fit $g_{\rho \gamma}$ and then BR$(\rho^0 \ra e^+ e^-)$, which was
previously simply fixed at its PDG \cite{PDG} value
{\footnote {This parameter was
taken as fixed in Refs.~\cite{benayoun,benayoun2}
because the authors believed that the
PDG value for BR$(\rho^0 \ra e^+ e^-)$ relied  on $e^+ e^-$ data
{\it and} other data. As there is no other data than $e^+ e^-$ annihilations
for this branching fraction, it should be fitted.}}.
The unitarised version of the HLS model is close
to the phenomenological model M$_3$ of Ref.~\cite{benayoun2}.

We note that the non--resonant term introduced
by VMD1 and HLS models is constant at leading order
whereas M$_1$ of Refs.~\cite{benayoun,benayoun2} has a
$q^2$--dependent non--resonant piece. In this respect the WCCWZ model is
close to M$_1$.

\subsection{Phase of $F_\pi(q^2)$
and Phase of the $\pi \pi \ra \pi \pi$ Amplitude}\label{formfc}

\indent \indent
{From} general properties of field theory (mainly, unitarity and $T$--invariance),
it can be shown \cite{gasio} that for $s \equiv q^2$ real above
threshold, we have

\be
F_{\pi}(s)=\exp{[2i\delta_1^1]}~~F_{\pi}^*(s)~~~,
\label{formfc1}
\ee

\noindent
up to the first open inelastic threshold.
Using Eq. (\ref{unit1}) and Eq. (\ref{unit2}) -- which defines
the $N/D$ formalism--, this relation gives \cite{gasio}

\be
F_{\pi}(s)=\frac{1}{D(s)}~~,
\label{formfc2}
\ee

\noindent where $D(s)$ has been defined by its general properties
in the preceding section.  In order  that $F_{\pi}(0)=1$, the
$D$ function should fulfill{\footnote{ In phenomenological applications,
one could prefer requiring a condition on a physically accessible
invariant mass region as, for instance, the two--pion threshold, where
we have $F_{\pi}(4 m_{\pi}^2)=1.17 \pm 0.01$ from ChPT
estimates, as it will be seen below.}} $D(0)=1$. The functions $D$ appearing
here and in Eq (\ref{unit2}) can be chosen identical without any loss of
generality. It is obvious from Eq. (\ref{formfc2}) that $F_{\pi}(s)$
carries the same phase  as $1/D(s)$ and as the $\pi \pi \ra \pi \pi$
amplitude.

The identification of the function $D$ in Eq. (\ref{unit2}) with a
resonance propagator, namely the $\rho^0$ meson (see Eq. (\ref{unit3})),
would be motivated in the present
case{\footnote{Namely: $s < 1$ GeV$^2$, $I=l=1$
$\pi \pi \rightarrow \pi \pi$ partial wave.}} by the fact that the
amplitude is dominated by a single resonance (the $\rho^0$ meson
itself) \cite{estabrooks,baton,proto}.
This does not mean that higher mass resonances, which surely
exist \cite{barkov,aleph}, have a (strictly) zero contribution
in our mass range, but simply that their magnitude is negligible
compared to that of the $\rho$ meson. In this case, the Breit-Wigner
parametrisation is flexible enough in order to absorb these
small contributions into a correspondingly small
change of the parameter values with respect to their (unknown)
``true" values. This is also valid for a possible small non-resonant
hadronic contribution to the amplitude.

As far as hadronic contributions to
the pion form factor are concerned, there are also contributions
from the $\omega$ and $\phi$ mesons. Because of their small width, we can
add their contributions to $F_{\pi}(s)$; this will produce departures
from Eq. (\ref{formfc2}), however always in very limited invariant
mass interval. Moreover, they do not contribute to the
$I=l=1$ $\pi \pi$ phase shift for obvious reasons.

 Putting aside the question of $\omega$ and $\phi$ mesons
for the reasons just given, it remains  to recall
that there is a direct $\gamma\pi\pi$
contribution to the form factor  present in the models
VMD1, HLS, and WCCWZ, whereas VMD2 has no such contribution.
These additional direct contributions, which are constant for
VMD1 and HLS models,  modify  the  total
phase near threshold where the hadronic ($\rho^0$) contribution is small
in magnitude. Therefore,  it is
appropriate to take the full phase of $F_{\pi}(s)$,
rather than the phase of the $\rho^0$ propagator only, which  allows
us to  extract the exact behaviour of the $\pi \pi \ra \pi \pi$
phase ($\delta_1^1$) in the threshold region.

\section{Simultaneous Fits of $e^+e^- \ra \pi^+ \pi^-$ and $\mu^+ \mu^-$ Data}
\label{fitcond}

Fitting the $e^+e^-$ data from threshold to about 1 GeV involves
three well-known resonances, $\rho^0$, $\omega$ and $\phi$. As the last
two are narrow, their parametrisation
is relatively simple. But, due to its broadness, the $\rho^0$
meson has given rise to long standing problems of parametrisation (see
Refs.~\cite{benayoun,PRoos,GS} and previous references
quoted therein). Moreover, as we have mentioned,
one can ask whether experimental data require the existence of
a non--resonant $\gamma\pi^+ \pi^-$ coupling.
The conclusion of Refs.~\cite{benayoun,benayoun2} is that
 data on $F_\pi(q^2)$ alone are insufficient to answer this question.
This is also relevant to the test of QCD
proposed by Chanowitz \cite{chan1,chan2}.

In addition to the mass and width, in the context of the class
of models having widths of the form given in Eq.~(\ref{mesnwid}), one requires

 only
one additional parameter, $\lambda$, to define the $\rho$ shape.
The resulting
fit turns out to depend not only on the
non-resonant
coupling, but also on whether unitarisation is used or not.
One approach \cite{IS} to this problem is to
perform a simultaneous fit of all $e^+ e^- \ra \pi^+ \pi^-$ data \cite{barkov}
and $e^+ e^- \ra \mu^+ \mu^-$ data \cite{shwartz}. Indeed, if the data are
precise enough, we could see a non--resonant coupling  in $e^+ e^- \ra
\pi^+ \pi^-$, which will (of course) be small in
$e^+ e^- \ra \mu^+ \mu^-$.  Therefore, from first
principles, a simultaneous fit to both data
allows us to decouple the $\rho$ from  any non--resonant
$\gamma \pi^+ \pi^-$ coupling. Naturally, to be of any use in this,
the \emuon data would have to be very good.
Until recently relatively precise
measurements were available only for the region
around the $\phi$ mass \cite{orsay,kurdadze}. However, a new
data set \cite{shwartz} collected by the {\sc olya}
collaboration is available and covers a large invariant mass interval
from 0.65 GeV up to 1.4 GeV. We shall see shortly whether
it is precise enough to constrain the $\rho$ parameters.
Thus, the data sets which will be used for our fits are those
collected by {\sc dm1}, {\sc olya} and {\sc cmd} which are
tabulated in \cite{barkov} (for \ep) and only the {\sc olya} data
of \cite{shwartz} for \emuon; these  data sets do not cover the
$\phi$ peak region.

\section{Results of $e^+e^-$ data analysis}
\label{fitee}

In all of the previously described models, except for WCCWZ, the fit to
$e^+ e^- \ra \pi^+ \pi^-$ and $e^+ e^- \ra \mu^+ \mu^-$  data depends on
only five parameters.
The first four are the three $\rho$ meson parameters
(${m}_{\rho}$, $\Gamma_{\rho}$ and $\lambda$) and the
Orsay phase ($\phi$).
These are common to both VMD and HLS. In the VMD models
an additional parameter, $\epsilon$, has been introduced
in order to account for universality violation (see Eqs.~(\ref{epsy})
and (\ref{1vgam})),
while in the HLS model this parameter is replaced by $a$ (see
Eq.~(\ref{epsa})).
These last parameters allow us to fit the branching fraction
$\rho^0 \ra e^+ e^-$ within each model in a consistent way.
The WCCWZ model
depends on one additional parameter, $b$, which permits
a more flexible form for the non--resonant contribution, as
compared with the VMD1 or HLS models. Let us note that
introducing \emuon in our fit procedure together with \ep
does not require further free parameters.

Generally speaking, the parameter named $A$ in the VMD models above determines
the branching ratio Br$(\omega \ra \pi^+ \pi^-)$ and should be set free since
its value is strongly influenced by the data on \ep we are fitting. However,
in order to minimise the number of fit parameters at the stage when different
models are still considered, we fix its value from the corresponding world
average value \cite{PDG} of Br$(\omega \ra \pi^+ \pi^-)$.
We shall set $A$ free for our last fit, in order to get an optimum estimate of
Br$(\omega \ra \pi^+ \pi^-)$; this will be done only for the model
which survives all selection criteria.

Finally the fits have been performed for both the standard VMD, HLS
and WCCWZ models and their unitarised versions, for both
$e^+ e^- \ra \pi^+ \pi^-$ data alone and simultaneously with
$e^+ e^- \ra \mu^+ \mu^-$.
As all measurements in the region of the $\phi$ resonance
for each of these final states are not published
as cross sections \cite{orsay,kurdadze}, they are not taken into
account in our fits. When fitting the data, we take into account
the statistical errors given in \cite{barkov} for each \ep
data set. {\sc dm1} and {\sc cmd} claim negligible systematic errors
(2.2\% for {\sc dm1} and 2\% for {\sc cmd}, while the statistical errors
are typically 6\% or greater); these errors can thus
be neglected with respect to the quoted statistical errors. {\sc olya}
claims smaller statistical errors but larger systematic errors:
these two errors have comparable magnitudes from the $\rho^0$ peak
to the $\phi$ mass. We do not expect a dramatic influence from neglecting
these systematic errors, except that this would somewhat increase
the $\chi^2$ value at minimum and hence worsen slightly the fit quality.

The results are displayed in Table \ref{table1} (non--unitarised models) and
in Table \ref{table2} (unitarised models).
We show the fitted parameters in the upper section of each
table, while in the lower part we
provide the corresponding values for derived parameters of relevance.

\begin{table}[htb]
\begin{center}
\begin{tabular}{|c||c|c|c|c|}
\hline\hline
\hhhb {Parameter} & {VMD1 } & VMD2 & HLS & PDG  \\
\hline\hline
\hhhb
$\epsilon$ & $0.210^{+0.016}_{-0.018}$ & $0.163^{+0.007}_{-0.008}$ & --
& -- \\
\hhha
HLS $a$ & -- & -- & $2.399^{+0.028}_{-0.012}$ & --\\
\hhha
$m_{\rho}$ (MeV)      & $751.4^{+3.7}_{-2.8}$ & 776.74$\pm$2.2
&$755.1^{+4.9}_{-2.8}$ &  $769.1 \pm 0.9$ \\
\hhha
$\Gamma_{\rho}$ (MeV) &$146.0\pm 2.2$ & $145.10^{+2.1}_{-1.9}$ &
$143.32^{+1.8}_{-2.0}$ &151.0 $\pm$ 2.0\\
\hhha
$\phi$ (degrees)&$113.8^{+5.2}_{-6.9}$ &$106.3\pm 4.5$&$120.6^{+4.6}_{-5.7}$
&
-- \\
\hhha
$\lambda$ & $4.49^{+0.35}_{-0.43}$ & $1.61^{+0.34}_{-0.31}$ &
$3.92^{+0.34}_{-0.72}$& -- \\
\hhha
$\chi^2/{\rm dof}$ ($\pi \pi$) & 63/77 &148/77 &64/77 & -- \\
\hhha
$\chi^2/{\rm dof}$ ($\pi \pi + \mu \mu$) &105/115 &194/115 &108/115 & -- \\
\hline\hline
\hhha
$g_{\rho \gamma}$ (GeV$^{2}$) & $0.113 \pm 0.003$ & $0.119 \pm 0.002$ &
$0.115
\pm 0.003$ &0.120 $\pm$ 0.003\\
\hhha
$g_{\omega \gamma}$ (GeV$^{2}$)& -- & -- & -- & 0.036 $\pm$ 0.001\\
\hhha
$g_{\rho\pi\pi}^2/4\pi$ & $2.91 \pm 0.05$ & $2.76 \pm 0.03$  & $2.84
\pm 0.05$ &--\\
\hhha
$\Gamma(\rho \ra e^+e^-)$ (keV) & $6.72^{+0.38}_{-0.33}$& $6.74 \pm 0.20$ &
$6.34^{+0.45}_{-0.27}$& $6.77\pm 0.32$\\
\hline\hline
\end{tabular}
\parbox{130mm}{\caption{Results from fits to $F_\pi(q^2)$ and
$F_\mu(q^2)$ without unitarisation for the VMD1, VMD2,
and HLS models.  Shown for comparison are the Particle Data Group
quoted values \protect\cite{PDG}.
}
\label{table1}}
\end{center}
\end{table}

\begin{table}[htb]
\begin{center}
\begin{tabular}{|c||c|c|c|c|}
\hline\hline
\hhhb {Parameter} & {VMD1 } & VMD2 & HLS & WCCWZ  \\
\hline\hline
\hhhb
$\epsilon$ & $0.167 \pm 0.008$ & $0.215\pm 0.010$ & -- &
$0.142\pm 0.014$ \\
\hhha
HLS $a$ & -- & -- & $2.364 \pm 0.015$ & --\\
\hhha
WCCWZ $b$ (GeV$^{-2}$) & -- & -- & -- & $-0.319^{+0.139}_{-0.117}$\\
\hhha
$m_{\rho}$ (MeV)      & $774.67 \pm 0.65$ & $780.37 \pm 0.65 $
&$775.15 \pm 0.65$ &  $770.89^{+1.75}_{-1.51}$ \\
\hhha
$\Gamma_{\rho}$ (MeV) &$147.11\pm 1.60$ & $155.44 \pm 1.95$ & $147.67
\pm 1.47 $& $140.6^{+3.2}_{-2.9}$\\
\hhha
$\phi$ (degrees)&$94.7\pm 4.3$ &$98.8\pm 4.4$&$105.1 \pm 4.3$&
$101.7\pm 5.3$ \\
\hhha
$\lambda$ & $1.038^{+0.080}_{-0.085}$ & $0.567 \pm 0.055$
&$1.056 \pm 0.042$ & $1.623^{+0.231}_{-0.269}$ \\
\hhha
$\chi^2/{\rm dof}$ ($\pi \pi$) & 65/77 &81/77 &65/77 & 61/76 \\
\hhha
$\chi^2/{\rm dof}$ ($\pi \pi + \mu \mu$) &104/115 &128/115 &111/115 &
103/114\\
\hline\hline
\hhha
$g_{\rho \gamma}$ (GeV$^{2}$) & $0.118 \pm 0.001$ & $0.122 \pm 0.001$ &
$0.114\pm 0.001$ & $0.133\pm 0.007$\\
\hhha
$g_{\rho\pi\pi}^2/4\pi$ & $2.81 \pm 0.03$ & $2.94 \pm 0.04$  & $3.08
\pm 0.03$ & $2.08\pm 0.04$\\
\hhha
$\Gamma(\rho \ra e^+e^-)$ (keV) & $6.70 \pm 0.11$& $6.99 \pm 0.16$ &
$6.23 \pm 0.11$& $8.62\pm 0.46$\\
\hline\hline
\end{tabular}
\parbox{130mm}{\caption{Results from fits to \protect{$F_\pi(q^2)$}  and
\protect{$F_\mu(q^2)$} for the unitarised VMD1, VMD2, HLS and
WCCWZ models.}
\label{table2}}
\end{center}
\end{table}

\begin{figure}[htb]
  \centering{\
     \epsfig{angle=0,figure=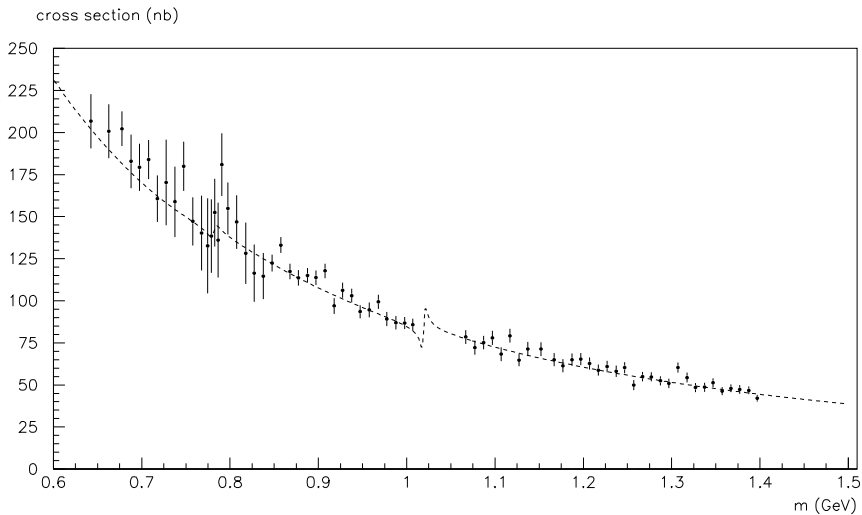,width=0.8\linewidth}
               }
\parbox{130mm}{\caption
{The $e^+e^- \ra \mu^+ \mu^-$ cross section. The data are shown together
with the fit from unitarised VMD2.  Other models
give similar descriptions.}
\label{figmu} }
\end{figure}

We find that, disappointingly, the new muon data\cite{shwartz} places
no practical constraint on the $\rho$ parameters extracted from $F_\pi(q^2)$.
Indeed, the central values for the fit parameters at minimum $\chi^2$ are
practically the same when fitting only $\pi \pi$ than $\pi \pi +\mu \mu$ data,
the errors become slightly larger in the second case (because of the
magnitude of the errors in the resonance region with the $\mu \mu$ final state).
The errors quoted in Tables  \ref{table1} and \ref{table2} are those obtained
when fitting only $\pi \pi$ data up to 1 GeV/c.
Thus the single existing data set on the muon final state is not precise
enough for our purposes.  To improve the situation we require more
accurate muon data below $\simeq 1$ GeV/c.  This is illustrated
by Fig.~\ref{figmu} which shows the muon data together with one of the best
fits (namely unitarised VMD2).

\begin{figure}[htb]
  \centering{\
     \epsfig{angle=0,figure=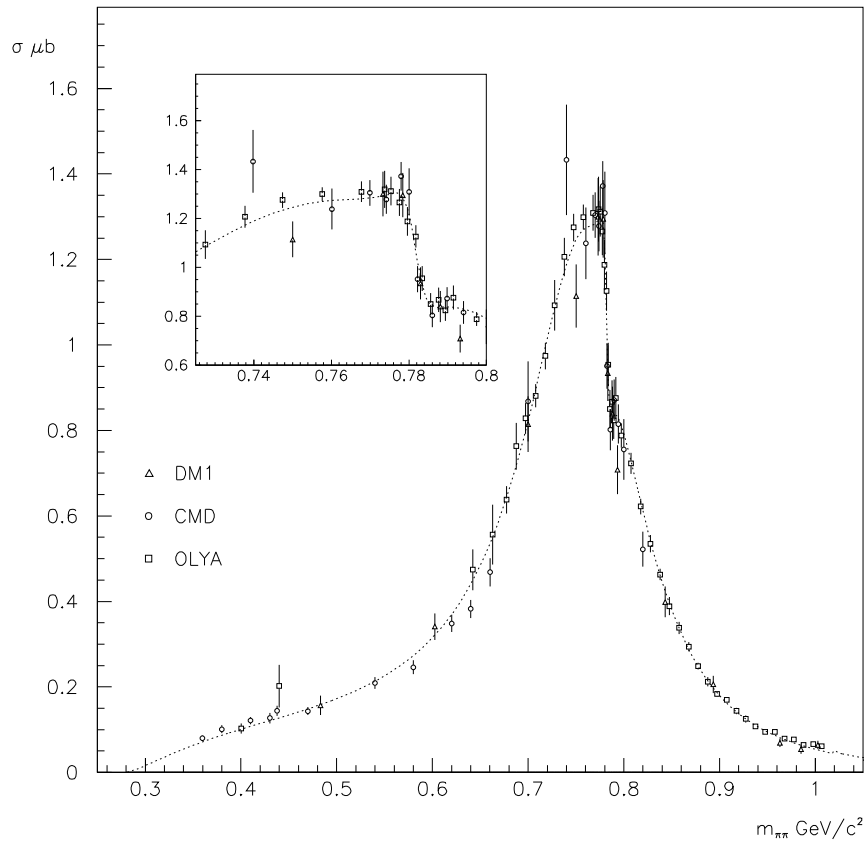,width=0.8\linewidth}
               }
\parbox{130mm}{\caption
{$e^+e^- \ra \pi^+ \pi^-$ cross section. The data are shown together
with the fit curve obtained using VMD2 not unitarised.}
\label{figvmd2}}
\end{figure}

\begin{figure}[ht]
  \centering{\
     \epsfig{angle=0,figure=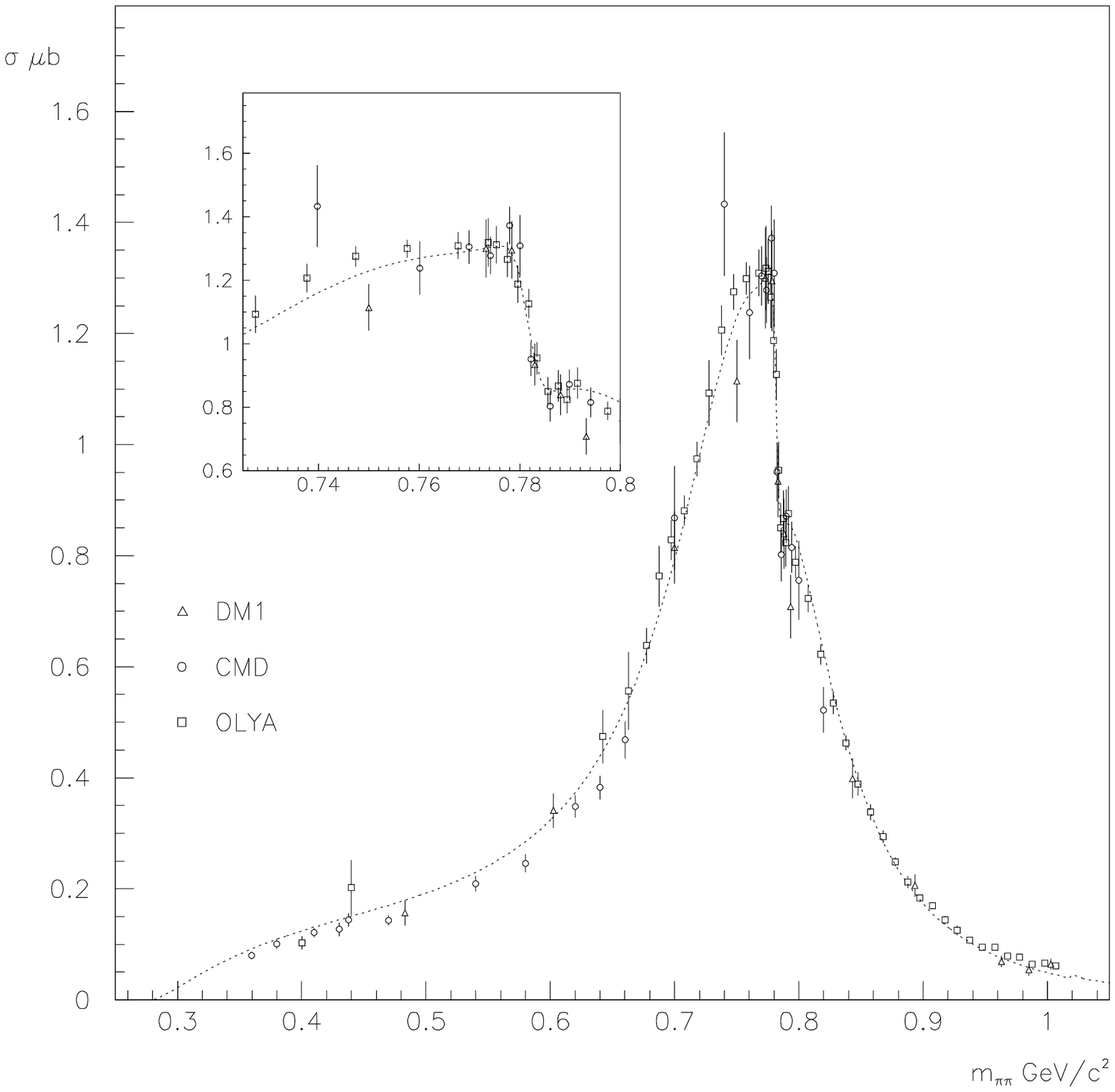,width=0.8\linewidth}
               }
\parbox{130mm}{\caption
{$e^+e^- \ra \pi^+ \pi^-$ cross section. The data are shown together
with the fitted curve obtained using unitarised VMD2.}
\label{figvmd2u}  }
\end{figure}

Another striking conclusion is that it is generally possible to achieve
a very good fit to the pion data whichever model is used, unitarised
or not; the single exception to this being non-unitarised VMD2
(which is the usual model for $\rho$ physics).  Correspondingly,
the significant model dependence of the extracted $\rho$ mass should be noted.
We do not give results in case of the non--unitarised WCCWZ model, as in this
case the solution converges to $b\simeq0$ and then coincides with VMD1.
We do not  present pion data curves for VMD1 and HLS as they show results
indistinguishable from Fig.~4 of Ref.~\cite{benayoun} which is probably the
best possible fit ($\chi^2$/dof=61/77).
Fig.~\ref{figvmd2} shows the fit obtained using non-unitarised VMD2; one
clearly sees that the model fails to describe the mass region below the
$\omega$ mass{\footnote{The region which strongly worsens the $\chi^2$
given in Table \ref{table1} is the mass interval from 400 to 600 MeV.}},
while the region from
the $\omega$ to the $\phi$ mass is quite correctly
reproduced. It is doubtful that taking into account higher mass
mesons could cure this problem. On the other hand, Fig.~\ref{figvmd2u} shows
the fit obtained using unitarised VMD2 which is of good quality throughout
the mass range. Its quality is better than that of M$_2$ (see
Fig.~3 of Ref.~\cite{benayoun}) simply because
$g_{\rho \gamma}$ (actually, $\epsilon$) is treated here as a
free parameter, as it should be.

For the non-unitarised  HLS model,
the $\rho$ mass and widths appear similar to those obtained
in the  fit by Bernicha {\it et al.} \cite{bernicha},
(${m}_\rho=757.5\pm1.5$ MeV and $142.5\pm3.5$ MeV respectively),
which used propagators with constant
widths defined in terms of the complex pole locations.
The reason can be traced back to the fact that the HLS model mimics
quite well a Laurent expansion  when there is no
mass dependence in the numerator of the $\rho$ contribution
as in Eq.~(\ref{fbando}).
The unitarised fits (shown in Table \ref{table2}), however,
all show much higher $\rho$ masses than do the non--unitarised fits.

The values obtained for the widths
$\Gamma(\rho \ra \pi^+ \pi^-)$ and $\Gamma(\rho \ra e^+ e^-)$ and the
Orsay phase, $\phi$, are in the expected range. The  value
for $g_{\rho \gamma}$ is approximately three times that of $g_{\omega \gamma}$
as expected from SU(3) symmetry and ideal mixing
\cite{DM}, and $g_{\rho \pi \pi}^2/4 \pi$ is always at the expected
value (about 3) except for the  unitarised WCCWZ model which provides
a much smaller value.
Moreover, the smallness of the statistical error on $g_{\rho \gamma}$
should be noted,{\it  i.e.}, the uncertainty is essentially all due to
model dependence.

The value for $\lambda$ deviates from 1 [see the discussion surrounding
Eq.~(\ref{mesnwid})] quite substantially for the standard VMD1 and HLS
models, although the unitarised versions
return a more standard value. As  $\lambda$ is {\it a priori} an effective
parameter
which accounts for higher order effects in the perturbation expansion
and/or dynamical left--hand singularities, this is not yet  grounds
for rejection of the non--unitarised versions of the models.
On the other hand, unitarised VMD2 provides,
as expected, a value for $\lambda$ close to that found using M$_2$
in Ref.~\cite{benayoun}. We also note that universality ($\epsilon=0$ or
$a=2$) is broken significantly in all fits at the level of $\simeq 20 \%$.
Finally, the effective
parameter $b$ introduced in the  unitarised
WCCWZ model stays relatively small
and allows one to obtain a more conventional mass for the $\rho^0$
meson\cite{PDG}.

It should be noted that, fixing $\lambda$ to 1 in the non--unitarised
versions of VMD1 and HLS, we essentially recover the solutions given
in Table \ref{table2}, as remarked in subsection \ref{unit};
the $\chi^2$ value is practically as good. The
situation is completely different for VMD2, for which  $\lambda$
is found significantly different from 1. As there is no general
requirement based on unitarity and analyticity which constrains the value of
$\lambda$, nothing can be concluded from this observation. It is however
interesting that unitarisation of the VMD1 and HLS models returns
$\lambda \simeq 1$; comparing for instance the results for the HLS
model given in Table \ref{table2} with the corresponding results for
model M$_3$ in Ref.\ \cite{benayoun2}, clearly shows that this is
a consequence of having released $\Gamma(\rho^0 \ra e^+e^-)$ in the fits.

As far as one relies only on the statistical quality of fits  for the cross
section $e^+ e^- \ra \pi^+ \pi^-$, the existing data do not allow one
to determine the most suitable way to implement vector meson dominance,
except  to discard the non--unitarised version of VMD2 which is
clearly  disfavoured by the data.
On the other hand, the possible values for the $\rho^0$ mass
cover a wide mass range: from 750 MeV to 780 MeV.
The single firm conclusion which can be drawn from the above
analysis is that, whichever is the VDM parametrisation
chosen, unitarised or not, one always observes a small but statistically
significant signal of universality violation: $\epsilon \simeq 0.20$
(instead of 0) or $a \simeq 2.4$ (instead of 2).

The question, therefore, remains as to whether it is possible to find
other criteria to distinguish between the various ways of building
effective Lagrangians involving the vector mesons.

\section {Comparison with $\pi \pi$ Phase shift Analysis}
\label{phaseshift}

 Until  now, we have focussed on the various ways
of implementing the VMD assumption when
describing the $e^+ e^- \ra \pi^+ \pi^-$ cross section.
As is clear from Section \ \ref{models}, the way to
express VMD is not unique. A representative set of possible models has been
given in the previous section.  We say ``representative'' in the sense that
it is difficult to imagine a class of models very
different from the four models studied in both unitarised and
non--unitarised forms.

The previous section illustrates that essentially all of the known
formulations of VMD are able to provide a quite satisfactory description
of the $e^+ e^-$ data.  This was a conclusion previously reached
in Refs.\ \cite{benayoun,benayoun2} with models referred to in that
work as M$_1$, M$_2$ and M$_3$.  As the $e^+ e^- \ra \pi^+ \pi^-$
cross section is not sufficient to differentiate between
the various realisations of
VMD, comparison with other data and/or information is helpful.

The $e^+ e^- \ra \pi^+ \pi^-$ cross section depends only on
$|F_\pi(q^2)|$, i.e., the magnitude of $F_\pi$.  We have
at our disposal a number of models which give equally good fits to this
cross-section.  Hence we can now examine the ability of each of our
versions of the VMD hypothesis to reproduce the $I=l=1$ $\pi\pi$
phase shift data, as we have $\arg{[F_\pi(q^2)]} \equiv \delta_1^1$, as
noticed in Section \ref{formfc}.

 As previously noted the resonant part of the
form factor also contains in general some small isospin zero contributions
({\it e.g.}, from the $\omega$ and $\phi$).
Below 1 GeV, the dominant isospin 1 contribution
to strong interactions is the $\rho^0$. This was the conclusion previously
reached when analysing $\pi \pi$ data \cite{estabrooks,baton,proto}.
Moreover, it has been shown \cite{aleph} recently
in $\tau$ decay to pion pairs, that the influence of possible
higher mass vector isovector mesons ($\rho(1450)$ and $\rho(1700)$)
is negligible below $\simeq 1$ GeV. As $\tau$ decays
and $e^+ e^-$ annihilation are connected through CVC \cite{tsai}, which
has been shown in \cite{simref1}, the conclusion of {\sc aleph}
gives support to neglecting these states in our
invariant mass range.

 Thus, there is an apparent elasticity of the $P$--wave isovector
$\pi\pi$ scattering up to $\simeq 900 \div 1000$ MeV. Then ,
we can limit the relevant  resonant contribution to the
$I=l=1$ $\pi \pi$ phase shift to the single $\rho$
meson propagator.  For VMD2, this means that the
opposite of the $\rho^0$-propagator phase coincides with the
phase $\delta^1_1$ of the $I=l=1$ $\pi \pi$ partial wave
(see Eq.\ (\ref{unit5})). When dealing with all other models,
there is a  nonresonant  direct $\gamma\pi\pi$
contribution to the pion form
factor which also contributes.
It is the concern of the present section to
check to which extent, the phase of the form factors defined in
Section \ref{models} (together with the numerical values
obtained from fit to \ep and listed in
Tables \ref{table1} and \ref{table2}) gives an appropriate
description of  $\delta^1_1$, as it is determined and/or measured
in $\pi \pi$ scattering.

$\pi \pi$ scattering is an important physical process,
where the basic concepts of S--matrix theory can be applied
with some precision; these basic requirements include unitarity, analyticity
and crossing symmetry.  In the case of the $\pi \pi$ system, crossing
symmetry puts stringent constraints on the partial waves
by mean of the Roy equations\cite{roy} or dispersion relations.
Methods based on these general principles are well known
(see, {\it e.g.}, Ref.\ \cite{basdevant} and
references therein) and have been applied \cite{petersen} in order to
fit and/or reconstruct the $I=l=1$ $\pi \pi$ phase shift
$\delta^1_1$ at low energy ({\it i.e.} from threshold up to about 1 GeV).
These methods and most of their results are presently
encompassed by ChPT \cite{gasser1,gasser2,knecht,stern}.

Therefore it is of some interest to compare the phase shifts
predicted from  the versions of VMD obtained after fitting to the
$e^+ e^- \ra \pi^+ \pi^-$ data  to the values of $\delta^1_1$
as tabulated by Ref.~\cite{petersen} (see their Table 1).
In this way, it is possible to check the consistency
of the information deduced assuming each VMD formulation
with the results of Ref. \cite{petersen} which were derived under
completely independent assumptions (namely, unitarity, analyticity
and crossing symmetry). One should note that our model errors (those
produced by the errors on the fitted parameters) have a small
effect on the predicted phase; indeed, the accuracy on the predicted
phase is 1.5\% at 600 MeV and 0.6\% at 800 MeV.
{From} a practical point of view, Ref.\ \cite{petersen} does not quote
errors. Therefore, we shall  unfortunately not be able to express this
comparison in terms of statistical quality factors.

\begin{figure}[ht]
  \centering{\
     \epsfig{angle=0,figure=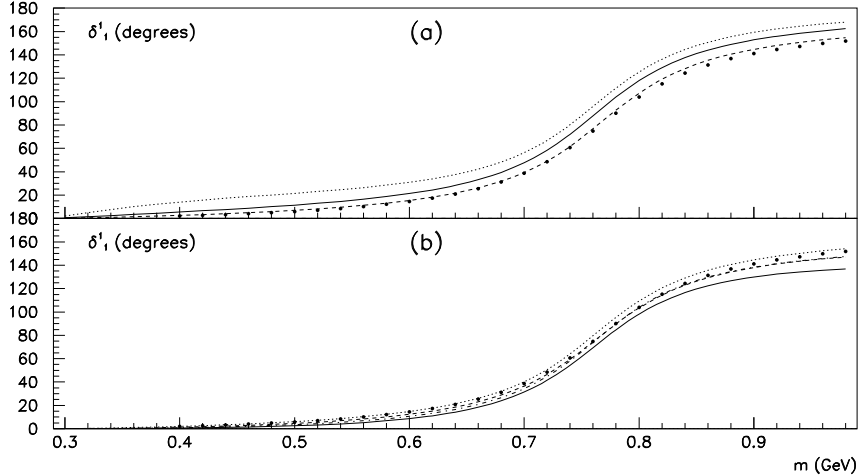,width=0.8\linewidth}
               }
\parbox{130mm}{\caption
{$\pi^+ \pi^-$  phase shift $\delta^1_1$. In both
figures, the points are those from Ref.\ \protect\cite{petersen}.
In (a) curves displayed correspond to non--unitarised
models (parameter values taken from Table \protect\ref{table1});
full line is VMD1, dotted line is HLS and dashed line is VMD2.
In (b) the curves displayed correspond to unitarised models
(parameter values taken from Table \protect\ref{table2}); the
uppermost curve (dotted) is HLS, the lowest curve (full line)
is VMD1; in between, the two curves corresponding to VMD2 and
WCCWZ  almost superimpose at this scale.}
\label{figadel4}  }
\end{figure}
\begin{figure}[ht]
  \centering{\
     \epsfig{angle=0,figure=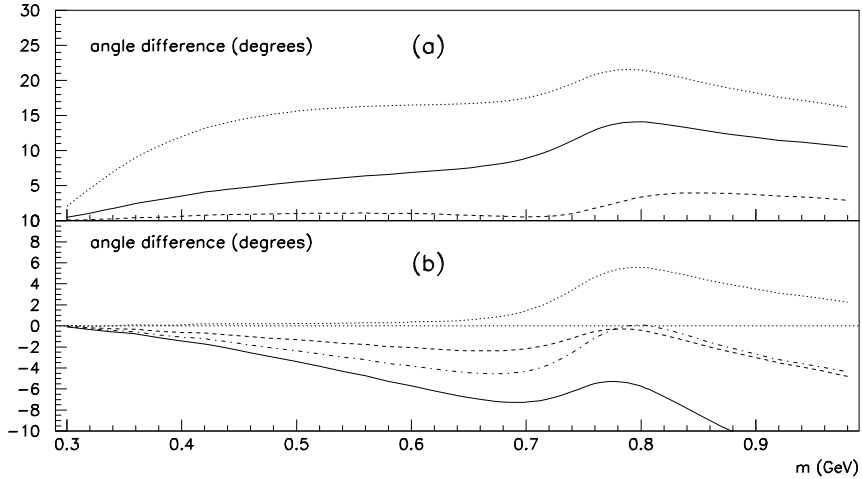,width=0.8\linewidth}
               }
\parbox{130mm}{\caption
{$\pi^+ \pi^-$  phase shift $\delta^1_1$.
 The plots display the {\it difference} between the phase
 shift of Ref.\ \protect\cite{petersen} and our phase shifts.
In (a) curves displayed correspond to non--unitarised
models (parameter values taken from Table \protect\ref{table1});
full line is VMD1, dotted line is HLS and dashed line is VMD2.
In (b) the curves displayed correspond to unitarised models
(parameter values taken from Table \protect\ref{table2}); uppermost
curve (dotted) is HLS, lowest curve (full line) is VMD1 and
middle curves are (dashed line) VMD2 and (dashed--dotted) WCCWZ .}
\label{figadel3}  }
\end{figure}

In Fig.~\ref{figadel4}, we have plotted the curves predicted
from our analysis and the phases from Ref.\ \cite{petersen}. The
non--unitarised versions of VMD1 and HLS can clearly be rejected
as failing to support the identification of the $\delta^1_1$ phase
shift with the phase of $F_{\pi}(q^2)$.
Non--unitarised VMD2 looks in better agreement with expectation;
however, recall
that non--unitarised VMD2 is in poor agreement with $e^+ e^-$ data
(see Table \ref{table1} above). These remarks are even clearer
from Fig.~\ref{figadel3}, where the difference between the model
phases and the data point phase of \cite{petersen} is plotted.
 Hence, we conclude that all non--unitarised forms of VMD are
disfavored by the data.

Instead, Fig.~\ref{figadel4} exhibits remarkably
good agreement of all unitarised versions of VMD with the
data from Ref.\ \cite{petersen} over the full invariant mass
range (from the two--pion threshold up to 1 GeV). Taking into account
the simplicity of our expressions for $F_{\pi}(q^2)$,
this is already remarkable.
The difference between the various
predictions for the phase shift and the data of Ref.\ \cite{petersen}
is presented in Fig.~\ref{figadel3}(b) for the unitarised models.
Clearly, VMD1 fails to describe Froggatt--Petersen phase over
the whole invariant mass range, at a finer level of accuracy.
Both WCCWZ and VMD2 have a
systematic disagreement with the Froggatt--Petersen phase,
(it is larger at low energy for WCCWZ than for VMD2). Finally, the
HLS model exhibits a quite remarkable agreement with the expected phase
up to about 700 MeV, {\it i.e.} where the  Froggatt--Petersen
reconstruction is expected to be highly accurate; in this
region the distance to expectation grows slowly, up to only
0.5 degrees near 700 MeV.

Given that the {\it pole} location for the $\rho$ in the complex
$s$--plane is an input into the phase shift analysis{\footnote
{We thank C.D. Froggatt and J.L. Petersen for this information.}}
(set at a mass of 767 MeV and a width of 137 MeV and not allowed to
vary), the disagreement of about 6 degrees near the $\rho$
mass for the  HLS model should not be considered presently as particularly
significant and, correspondingly, the fact that the point where the phase
goes through $\pi/2$ in Ref.\ \cite{petersen} (780 MeV) is
close{\footnote {This is {\it not} an input in their method, but an output.
}} to the mass given by VMD2 (780.4 MeV) should also not be
regarded as especially remarkable{\footnote {In order for comparison at
this level of refinement to be meaningful, the Froggatt--Petersen
approach should first be carried out with a more accurate $\rho$ pole
location, those of Ref.~\cite{bernicha} for instance.}}.

Finally, the relative difficulty for our unitarised models
to match the phase above $800 \div 900$ MeV (see Figs.
\ref{figadel4}b and \ref{figadel3}b) can also
be partly attributed to higher vector mesons which definitely
exist \cite{barkov,aleph} and may indeed contribute at a small
level in this mass region. However, such additional contributions
should not be significant below the $\rho$ peak.
One should also notice that the phase shift of \cite{petersen}
is not expected to be as accurate in this invariant mass region as it
should be below the $\rho$ mass peak. For all these reasons, the
matching of phases in the region $800 \div 900$ MeV cannot be
considered as constraining as the matching in the region $300 \div 600$ MeV.

\begin{figure}[ht]
  \centering{\
     \epsfig{angle=0,figure=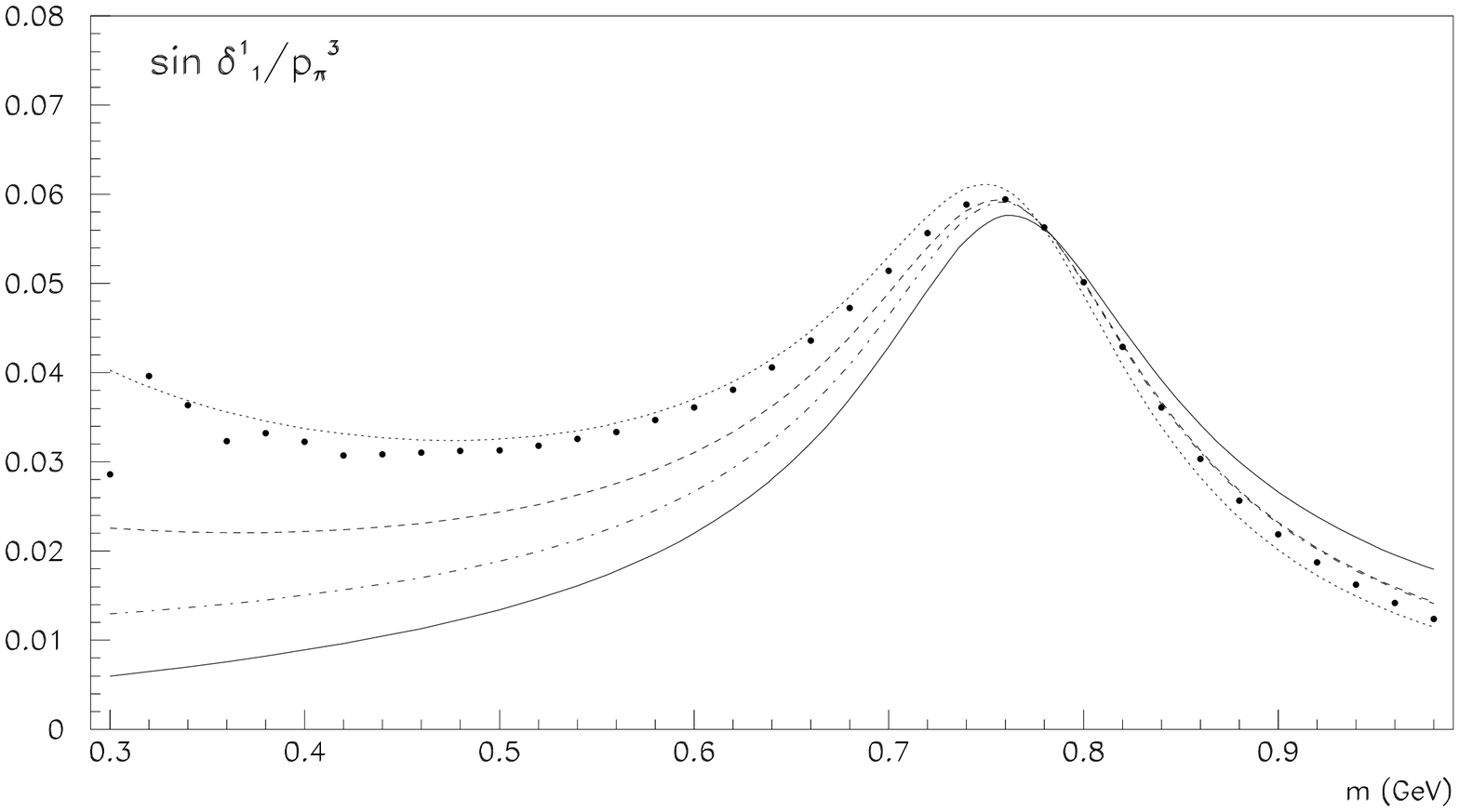,width=0.8\linewidth}
               }
\parbox{130mm}{\caption
{Function $\sin{\delta^1_1}/p_{\pi}^3$ deduced from $\pi^+ \pi^-$
phase shifts; the function is given in units of $m_{\pi}^{-3}$.
The dots correspond
to the points of Ref.\ \protect\cite{petersen}. Full line is VMD1,
dashed line is VMD2, dotted line is HLS (the best description
of expected phase) and dashed--dotted is WCCWZ.}
\label{figadel5}  }
\end{figure}

\begin{figure}[htb]
  \centering{\
     \epsfig{angle=0,figure=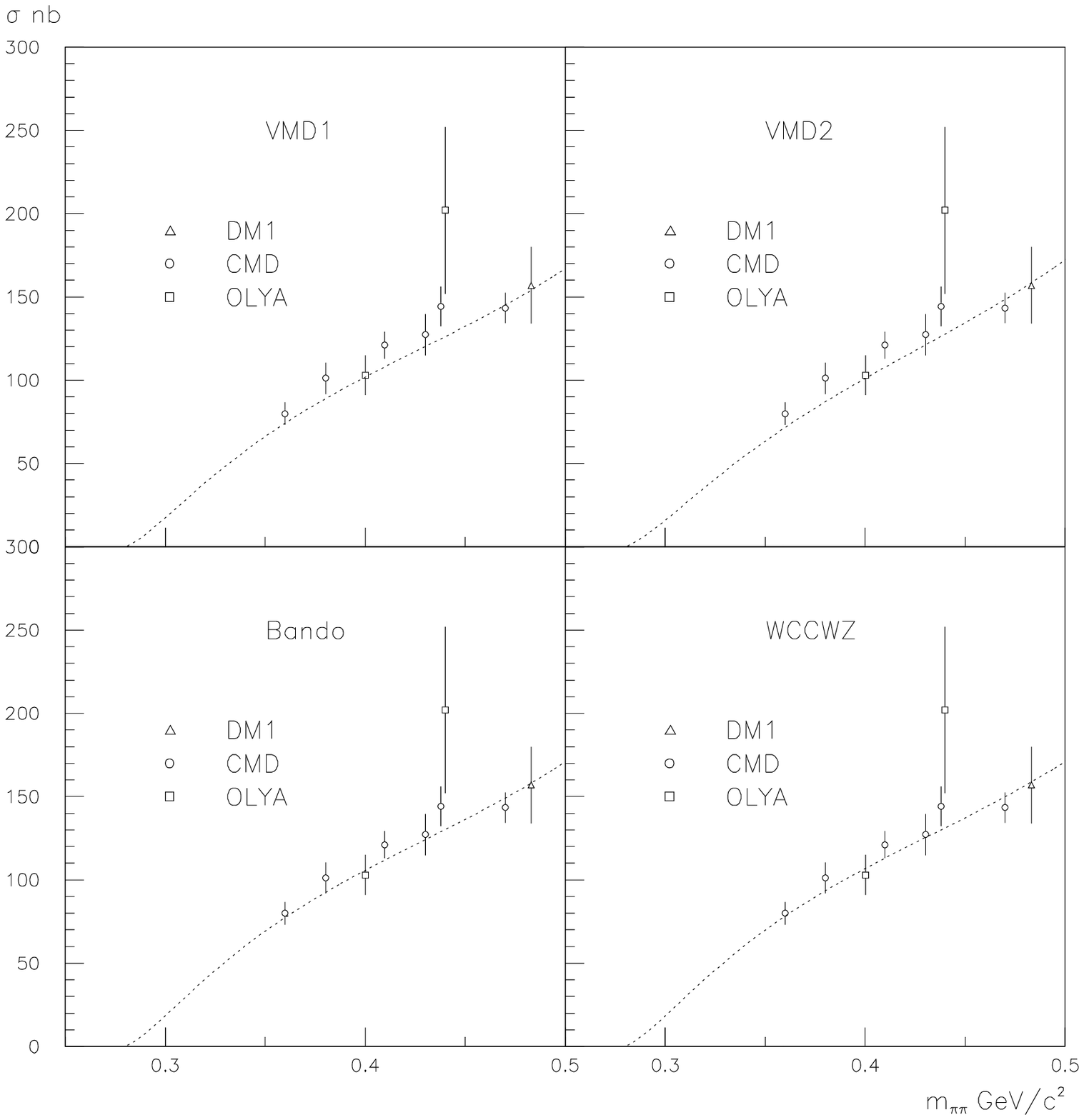,width=0.8\linewidth}
               }
\parbox{130mm}{\caption
{$e^+e^- \ra \pi^+ \pi^-$ cross section in nb with fits
using various unitarised VMD models. Data point and curves are
given from threshold to 500 MeV.
}
\label{figscatt}}
\end{figure}

Another way to carry on the comparison of the $F_{\pi}(q^2)$ phases
with the Froggatt--Petersen phase shift, is to
superimpose the  various model predictions for
$\sin{\delta^1_1}/p_{\pi}^3$.
This function is connected with the  isospin 1 $P$--wave
scattering length $a_1^1$ through:
\begin{equation}
a_1^1= \displaystyle \lim_{q^2 \ra 4 m_{\pi}^2}
\frac{\sin{\delta^1_1(q^2)}}{p_{\pi}^3(q^2)}
\label{scatlen1}
\end{equation}
and  has the  characteristic of highly magnifying differences due
to the phases in the low energy region (up to, say, 600 MeV).
The curves corresponding
to the various unitarised models are shown in Fig.~\ref{figadel5} together
with the points reconstructed using the phase as given in
Ref.\ \cite{petersen}.
 Note that erratic behaviour of the values deduced from
Ref.\ \cite{petersen} at the very beginning of the curve
is entirely due to rounding errors only{\footnote
{Ref.~\cite{petersen} gives for the phases 0.1$^{\circ}$,
0.4$^{\circ}$, 0.7$^{\circ}$ at respectively 300, 320 and
340 MeV. In order to get a smooth behaviour consistent with the rest
of the curve, it is enough to change these angle values to
respectively 0.14$^{\circ}$, 0.36$^{\circ}$, 0.66$^{\circ}$.
This remark illustrates the magnification effect produced
by the function in Eq.\ (\ref{scatlen1}).}}.
The general success of the HLS model in  matching
the low energy $\pi \pi$ phase data is especially remarkable.
In terms of angles, in the low energy region the distance to
the data points does not exceed a few  hundredths of a degree. Instead
VMD1 finds 0.02$^{\circ}$, WCCWZ 0.03$^{\circ}$ and  VMD2 0.06$^{\circ}$,
where the prediction is rather 0.1$^{\circ}$.
In Fig.~\ref{figscatt} we present, for all unitarised models,
the measured cross section for $e^+e^- \ra \pi^+ \pi^-$ and the previous fit
function behaviours in the low energy region.
 They are equally as good with the possible exception of VMD2.
{From} this, we conclude that the difference in ability to
describe the $\delta^1_1$ phase shift is due to the existence and
the magnitude of a direct $\gamma \pi^+ \pi^-$ coupling
within models when fitting $e^+ e^-$ data.

The success of the HLS model in
reproducing the correct phase values and behaviour is
remarkable.
This can be partly attributed
to the fact that its  $\gamma \pi \pi$ coupling is connected
with universality violation, while for VMD1, this term
is fixed in absolute magnitude to the pion charge.
Another reason is connected with the fact that in the HLS
model, $g_{\rho \gamma}$ is constant while it is proportional
to $q^2$ in VMD1 and WCCWZ. This actually produces a too large
drop of the resonant contribution to the phase at low invariant
mass which is not supported by the data, as it is illustrated
in Fig.~\ref{figadel5}.

On the other hand, the WCCWZ model, while differing from all other
models  by having a conventional $\rho$ mass and
an additional $q^2$ dependent term, does not match the phase shift
better than VMD1 or VMD2. We may therefore conclude that
the extra term in WCCWZ is not consistent with appropriate low-energy
behaviour when WCCWZ is constrained to fit
the broad energy range ($0.28$ to $1.0$ GeV)
rather than near threshold only.

\vspace{0.8cm}

\indent
Figs.~\ref{figadel4}, \ref{figadel5} and \ref{figscatt}
demonstrate that, within the class of models
considered, a good fit to {\it all} experimental data can only be
obtained by including an {\it explicit} direct $\gamma\pi^+\pi^-$
coupling term  in the model.  For the HLS model, the magnitude
of this direct coupling is quite well-fixed by data and is
approximately $-e/6$ (in contrast to the common form of VMD, VMD2,
for which such a term is absent).

Indeed, Fig.~\ref{figscatt}, clearly shows that the four unitarised
models fit practically as well the cross section \ep in the low energy
region. Therefore, their relative ability to account here for the phase
$\delta^1_1$ does not originate from sharp differences in
fitting \ep in the same region. The phase corresponding
to the HLS model matches accurately $\delta^1_1$ \cite{petersen}
from threshold to about 700 MeV, as illustrated by Figs.~\ref{figadel4}
and \ref{figadel5}; this phase coincides with that of
$F_{\pi}(q^2)$. Instead, VMD1 having a  $\gamma \pi^+ \pi^-$ coupling
of magnitude fixed to $e$, fails to describe accurately this low energy
region.  VMD2 exhibits a systematic disagreement, which
can be traced back to the lack of an explicit
$\gamma \pi \pi$ coupling, more important at low energies; it should
be remarked that  VMD2 is, however, the closest (after HLS) to the data
points;
this is related to the fact that VMD2 actually corresponds to the
HLS model with $a=2$ while the data prefer a slightly different value
($a \simeq 2.4$).

The behaviour
of WCCWZ with respect to HLS  clearly shows that data
considered from threshold to about the $\phi$ mass are more consistent
with  the existence of a constant nonresonant
coupling. Another interesting result is that, from
Table \ref{table2}, it is clear that we can perform the fits with
the HLS model fixing $\lambda =1$ without any  significant
change in the fit quality
or in the parameter values. The fact that $\lambda \simeq 1$ is preferred
by the data at the fit level is interesting in a number of regards; indeed,
$\lambda =1$ is
one of the most common ways to parametrise the $\rho$ width; it
is also the value obtained for the width for an interaction
term in the Lagrangian of the form \cite{klingl}
$\rho^{\mu} (\pi^+ \partial_{\mu} \pi^- - \pi^- \partial_{\mu} \pi^+)$
with constant dressed coupling.
We discuss in Section \ref{discussion}, whether a departure
of $\lambda$ from 1, even if not significant at the fit level
(see Table \ref{table2}), should be considered.
 In closing this section we note that no {\it non--unitarised} model
appropriately reproduces the $\pi \pi$ phase shift \cite{petersen},
and that, among all unitarised models we have examined, only the HLS model
of Bando {\it et al.}
agrees well with the $\pi \pi$ phase shift predictions \cite{petersen}.

\section{ChPT and constraints in the near-threshold region}
\label{threshold}

In the preceding sections, we have analysed various
 realisations of
VMD using $e^+e^-$ data and a $\pi \pi$ phase shift
analysis, over the full invariant mass range where data and/or information
is available.  It is also of interest to examine
in detail the consequences of the various VMD models for the
magnitude of $F_\pi$ in the near threshold region.

{From} a general perspective (such as that of ChPT, for example), the
VMD Lagrangians are  rather simple and
should not be expected a priori to be able to provide an accurate
representation of the full physics of QCD over the whole of the low-energy
regime  below $1\ {\rm GeV}^2$. In particular,
it is known that the effects of $\rho$ exchange first show up
in the usual Gasser and Leutwyler effective chiral Lagrangian,
${\cal L}_{\rm eff}$, (relevant
to processes involving only external $\pi$ legs\cite{gasser2,gl85a,gl85b})
as contributions to the fourth order low-energy constants (LEC's),
$L_k^r$, appearing in ${\cal L}_{\rm eff}$\cite{EGPdR}.  This means that,
very near threshold, other contributions, not generated by $\rho$
exchange, are also, in principle, present in $\pi\pi$
scattering and $F_\pi$.  The models VMD1, HLS and WCCWZ allow
for such a possibility, though in a form much simpler than would
necessarily be expected on general principles.  The fact that we
allow some phenomenological freedom in the treatment of the $s$-dependence
of the $\rho$ width, however, means that some of the more complicated
$s$-dependence of the non-resonant contribution may be able to be
reproduced by a slight adjustment of the parameter $\lambda$.  Such
a readjustment, of course, has no effect at the $\rho$ peak.

The physical spectral function near threshold in the vector
isovector channel, which is
measured experimentally in $e^+e^-\rightarrow \pi^+\pi^-$
 and in $\tau$ decays\cite{aleph}, is very
well described  near threshold by ChPT\cite{gl85a,gl85b,kg95}.
At next-to-leading (two-loop) order in the chiral expansion,
the ChPT expression for $F_\pi (q^2)$\cite{gm91}, however, contains
two new sixth order LEC's not constrained by other experimental data,
so an improved {\it prediction} is not possible at present.
However, one can use the {\it form} of the two-loop prediction\cite{gm91},
\begin{equation}
F_\pi (q^2) = 1+{\frac{1}{6}}\langle r^2\rangle_V^\pi q^2
+c_\pi q^4 +f^U_V(q^2/m_\pi^2)+{\cal O}(q^6),
\end{equation}
(where $f^U_V$ is a known function, whose explicit form may be found
in Ref.~\cite{cfu96})
to analyse the near-threshold data and obtain an optimal model-independent
fit for the parameters $\langle r^2\rangle_V^\pi$ and $c_\pi$.
The results are\cite{cfu96}
\begin{eqnarray}
\langle r^2\rangle_V^\pi &=&0.431\pm 0.020\pm 0.016\ {\rm fm}^2 \nonumber\\
c_\pi&=& 3.2\pm 0.5\pm 0.9\ {\rm GeV}^{-4}
\end{eqnarray}
where the first error is statistical and the second error theoretical.
Note that, although the data analysed is that of the NA7
collaboration\cite{NA7}, the statistical error is roughly a factor
of 2 larger than quoted in Ref.~\cite{NA7}.  According to the
authors of Ref.~\cite{cfu96} this larger error reflects the
presence of the additional fit parameter, $c_\pi$.  We will accept
the results of the analysis of Ref.~\cite{cfu96} since
the form of the fit function used, being obtained from ChPT, is the
most general one possible, compatible with QCD, to this order in
the chiral expansion.  As such, the analysis is as model-independent
as possible, to this order.  Using these results we obtain an
experimental value for the threshold value of $F_\pi$.  In quoting
this result below we have added all errors in quadrature.  Note
that the next-to-leading order contributions appearing in
this experimental fit are in agreement with those appearing in the
1-loop ChPT expression for $F_\pi (q^2)$ \cite{gl85b}.
This is of interest since,
in the case of the one-loop expression, the relevant LEC, $L_9^r$,
is also constrained by other experimental data.  Although it is
conventional to fix $L_9^r$ using $\langle r^2\rangle_V^\pi$ as
input, fixing it independently through a quantity like the
combination $r_A/h_V$ appearing in the amplitude for
the process $\pi^+\rightarrow e^+\nu_e e^+e^-$ leads to compatible
values.  As a result, the 1-loop expression for $F_\pi$ may
be thought of as a prediction.

We now compare the results of the various  unitarised
models, as fitted
to $e^+e^-$ cross-section data, for $F_\pi (4m_\pi^2)$, with
the experimental value, obtained as described
above{\footnote {There was a small change in the ``experimental'' number
above from the $f^U_V$ term in the general expansion.}}

\begin{equation}
\left \{
\begin{array}{lll}
\left[ F_\pi (4m_\pi^2)\right]_{\rm VMD1}&= & 1.174 \pm 0.001 \\[0.2cm]
\left[ F_\pi (4m_\pi^2)\right]_{\rm WCCWZ}&= & 1.149 \pm 0.010 \\[0.2cm]
\left[ F_\pi (4m_\pi^2)\right]_{\rm HLS}&= & 1.176 \pm 0.001 \\[0.2cm]
\left[ F_\pi (4m_\pi^2)\right]_{\rm VMD2}&= & 1.393 \pm 0.001 \\[0.2cm]
\left[ F_\pi (4m_\pi^2)\right]_{\rm Expt}&= & 1.17 \pm 0.01 \\[0.2cm]
\end{array}
\right.
\label{fth}
\end{equation}

We see that each of these unitarised VMD models is close to the ChPT result
with the exception of VMD2.

In order to quantify our comparison of the low energy phase behaviour we have
computed the scattering lengths $a_1^1$ following from the various
unitarised models. For this purpose,  Eq. (\ref{scatlen1}) has been
used with the corresponding phase of each $F_{\pi}$, following our remarks
in Section \ref{formfc}; the results are:

\begin{equation}
\left \{
\begin{array}{lll}
a_1^1(\mbox{VMD1})&= &  0.006 \pm 0.007 \\[0.2cm]
a_1^1(\mbox{VMD2})&= &  0.023 \pm 0.002 \\[0.2cm]
a_1^1(\mbox{HLS})&= & 0.043 \pm 0.003 \\[0.2cm]
a_1^1(\mbox{WCCWZ})&= & 0.013 \pm 0.002 \\[0.2cm]
\end{array}
\right.
\label{scatlen2}
\end{equation}
in units of $m_{\pi}^{-3}$. These values can be
compared with experimental results from $K_{e4}$ data \cite{nigels,rosselet}
using a Roy equation fit ($a_1^1 =0.038 \pm 0.002$).  The result of
Ref.~\cite{Gesh} ($a_1^1 =0.10$) disagrees with these results and
that treatment violates basic assumptions of ChPT
as illustrated in Ref.~ \cite{basdevant2}.
The Current Algebra prediction \cite{weinberg} is
$a_1^1 =0.030$ and the ChPT result
\cite{knecht} predicts $a_1^1 =0.037 \pm 0.001$
at the two loop order (at ${\cal O}(p^4)$).  A recent preliminary
ChPT calculation  \cite{bijnens} predicts that the
expected value for $a_1^1$ should increase at order ${\cal O}(p^6)$
and give a value in the range $0.038 \div 0.040$.
The range given for this prediction reflects the existing uncertainties in
the low energy constants (LEC's).

\vspace{0.8cm}

\indent
The question with the above estimates of the scattering
length $a_1^1$ is twofold : {\bf 1)} to what extent is the interpolation of
each model into the low energy regime and subsequent extraction of
$a_1^1$ meaningful and reliable?  {\bf 2)} is it possible to improve each
estimate?

Assuming that the interpolation to low energies is meaningful,
which seems legitimate in view of the preceding section
(at least for the HLS model),
the answer to the ``reliability'' issue is connected with the
fit quality for each case, which is well summarised by the
line $\chi^2/{\rm dof}(\pi \pi)$ in Table \ref{table2}. As noted earlier,
the best fit quality reached with the existing
$e^+ e^- \rightarrow \pi^+ \pi^-$ data \cite{barkov} corresponds to
$\chi^2/{\rm dof}=61/77$. This value was already achieved in
Refs.~\cite{benayoun,benayoun2} with their model M$_1$; we get
it here with the WCCWZ model which is a six parameter fit
while M$_1$ was a five parameter fit (as our present models
VMD1, VMD2, HLS). {From} a statistical point of view, a
$\chi^2/{\rm dof}=61/76$ is unlikely to be reliably improved
and should be considered close to perfection and
hence, $a_1^1(\mbox{WCCWZ})$ cannot be significantly improved.
Taking into account the systematic disagreement with the phase
of Ref.~\cite{petersen} revealed by Fig.~\ref{figadel5}, we cannot
consider this estimate as reliable.

Unitarised VMD2 gives a very good (even if not perfect) fit
  to the pion form factor ($\chi^2/{\rm dof}=81/77$);
however, from Fig.~\ref{figscatt}, it appears to have some difficulty
accommodating all data points as well as the three other VMD models
in the low energy region; this can be attributed to the lack
of a non--resonant term which is present in the other models.

Finally, VMD1 and the HLS model  provide both a near
optimum fit  ($\chi^2/{\rm dof}=65/77$) to the  pion
form factor, and so cannot be dramatically improved.
We thus believe that, numerically, the extrapolation
is reliable and cannot be significantly improved. We shall
nevertheless examine the consequences of fixing $\lambda=1$
in the next section.

To summarise this section, we can say that among the unitarised
versions of the VMD assumption, each of VMD1, WCCWZ and HLS provide a good
extrapolation for the magnitude of $F_\pi$ all the way down
to threshold, while only the  HLS model  provides additionally
a good extrapolation of the phase $\delta^1_1$.
{From} the discussion above, we can give our extracted value
for the scattering length, $a^1_1=0.043 \pm 0.003$, which compares
well with the previous measurement \cite{nigels,rosselet}
(their  separation is $1.4 \sigma$).

As a consequence of these observations we conclude that,
 of the models considered, only
the unitarised HLS model, with $a\simeq 2.4$,
provides a successful representation of $F_\pi$ in both
magnitude and phase which
is valid in the whole region from threshold to the $\phi$ mass.

\vspace{1cm}

\section{Discussion}
\label{discussion}

\indent
\indent The previous analysis of the $e^+e^-$ data \cite{barkov}
and  its comparison with the phase shift
$\delta^1_1$ of Froggatt--Petersen \cite{petersen}, lead us to
conclude that the VMD1, VMD2, WCCWZ and HLS unitarised models provide a  good
description of a very large set of experimental data, as illustrated
by Table \ref{table2} and by Fig.~\ref{figadel4}. However, a
finer check on the low energy behaviour (see Fig.~\ref{figadel5})
highly favors the HLS model over all other ways to formulate
the VMD assumption. This conclusion is enforced if one includes the
known  information in the near--threshold  region provided by
experiments and ChPT (values for $F_{\pi}(4 m_{\pi}^2)$ and $a_1^1$).

This agreement is reached by allowing the HLS  $a$
parameter to slighly depart from its universality value: $a \simeq 2.4$
instead of $a = 2$. Relying on this model, we find that universality
is violated  at a level of 20\%. As a direct consequence of this,
we observe a non--resonant  direct coupling
$\gamma \pi^+ \pi^-$,
with strength $-0.182 \pm 0.008$. The corresponding term in VMD1
is 1 (due to the pion electric charge), while it is 0 for the widely
accepted VMD2. Universality violation does not prevent the HLS model
fulfilling the constraint $F_{\pi}(0)=1$ in a quite natural way.

Even if the influence of this  direct
term is the largest near threshold, it contributes
a significant improvement up to 1 GeV, because of interferences.
This becomes even  clearer if one  notes
that VMD2 is qualitatively equivalent to the HLS model by
simply removing the constant term{\footnote {The role played by $a$
in $g_{\rho \gamma}$ in the HLS model is then transferred to
$\epsilon$ in VMD2.}}
in its expression for $F_{\pi}$ (see section \ref{models}). Therefore,
we  have the first suggestive
evidence for a non--zero  point--like $\gamma \pi^+ \pi^-$ coupling.

{ The origin of the shortcomings of the
VMD1, VMD2 and WCCWZ models are readily understood.
For VMD2, universality violation (required by the data) produces
an $F_\pi$ which, extrapolated to $q^2=0$, has
$F_\pi (0) -1\equiv \delta >0$.
Not surprisingly, near threshold then, the fitted VMD2 model
exceeds data by almost
exactly $\delta$.
For VMD1 and WCCWZ, the problem is the near-threshold phase.
In a conventional effective field theory formulation\cite{WCCWZ}, however,
the direct pion-photon coupling term in $F_\pi$
(unavoidably present) has a non-zero phase
via $\pi\pi$ rescattering.  The truncated version
represented by VMD1 and WCCWZ does not allow for this
possibility and
the models turn out to be insufficiently flexible to fully compensate for this
deficiency by an adjustment of the parameters of the $\rho$ contribution.}

An interesting issue is also the observed inequivalence of VMD1 and VMD2.
One may argue that this could be a shortcoming due to having neglected
mass dependence in the $\rho-\gamma$ coupling produced by loop effects.
Following Ref.~\cite{klingl}, we have checked that this is not the case,
mainly because universality violation is clearly preferred by the data.

 Another interesting issue is  the value found for $\lambda$ ($1.06 \pm 0.08$)
in the unitarised HLS model. The departure from 1 is not statistically
significant as far as fits of $e^+e^-$ data  alone are concerned and
hence the question of
whether one can set $\lambda=1$ should be considered.
 Because of unitarisation (see subsection \ref{unit}), there is no
reason why one could not observe departures from 1, which  could
effectively account
for neglected contributions ($t-$ and $u-$channel exchanges,
non-resonant strong interaction among pions, higher mass
resonances, etc...). The fit value
for $\lambda$  shows that these neglected effects are small enough
 not to spoil the analytical shape generally expected for the
$\rho^0$ mass--distribution as is obvious from Table \ref{table2}.
However, if all consequences of fixing $\lambda =1$ were acceptable,
this will allow us to provide a 4 parameter fit to the pion
form factor (the smallest possible set): 3 of these parameters
describe the $\rho$ mass and couplings to $\pi^+ \pi^-$ and $e^+ e^-$,
and the Orsay phase which is tightly connected with the omega
isospin violating contribution.

We have done the  fit with
the HLS model, fixing $\lambda=1$ in order to get the full set of
parameters needed to check accurately the near--threshold results of
the preceding section. We have also left free the parameter $A$ which
governs the magnitude of Br$(\omega \ra \pi^+ \pi^-)$, as it is
mainly influenced by \ep data.
As expected, the fit quality is
practically unchanged ($\chi^2/{\rm dof}=66/77$).
We have obtained for $F_{\pi}(4m_{\pi}^2) = 1.177\pm 0.001$,
in good agreement with the ChPT expectation
and the value reconstructed from other
experimental data \cite{cfu96,NA7}   ($1.17 \pm 0.01$).
We have also obtained $a^1_1=0.041 \pm 0.003$ using the full
phase of $F_{\pi}(q^2)$,  which is in better agreement
with the two--loop calculation of \cite{knecht} ($0.037 \pm 0.001$).
Moreover, our result is in fairly good
agreement with  \cite{basdevant2} which predicts
$a^1_1=0.040 \pm 0.003$, relying on Roy equation techniques.
This is consistent with the $\pi \pi$ $P$--wave being
nearly purely resonating with only a small background
from threshold to 900 MeV. It therefore appears legitimate
to fix $\lambda=1$ in our fitting procedure with the
unitarised HLS model.

We give as final results those obtained in this last
kind of fit with the HLS model \cite{bando}:

\begin{equation}
\left \{
\begin{array}{lll}
m_{\rho}(\mbox{MeV})&= & 775.1 \pm 0.7\\[0.2cm]
\Gamma_{\rho \ra \pi^+ \pi^-}(\mbox{MeV})&=
& 147.9 \pm 1.5\\[0.2cm]
 \Gamma_{\rho \ra e^+ e^-}(\mbox{keV})&=
&  6.3 \pm 0.1\\[0.2cm]
\rm{Br}(\omega \ra \pi^+ \pi^-)&=& (2.3 \pm 0.4) \% \\[0.2cm]
 F_\pi (4m_\pi^2)&= & 1.177 \pm 0.001\\[0.2cm]
a_1^1(m_{\pi}^{-3})&= & 0.041 \pm 0.003 \\[0.2cm]
\phi&= & 104.7^{\circ} \pm 4.1^{\circ}\\[0.2cm]
a(\rm{HLS})&= &2.37 \pm 0.02\\[0.2cm]
\lambda&= & 1 ~~(\rm{fixed})
\end{array}
\right.
\label{respdg}
\end{equation}

Concerning $F_\pi (4m_\pi^2)$ the value above is in
good agreement with ChPT predictions \cite{gl85b}
and the previously quoted experimental value.
Our value for $a_1^1$ compares well with the determination given in
 \cite{rosselet}, which relies on $K_{e4}$ data and the use of
Roy equations \cite{roy}, and with the ChPT calculation of \cite{knecht}
and the Roy equation result \cite{basdevant2}. Taking  into account the
uncertainties on the low energy constants (LEC's),
ChPT at order $p^6$ appears consistent \cite{bijnens} with $a_1^1=0.040$,
in good agreement with our extracted value.

 A final remark concerning the HLS model is of relevance.
As said above, it predicts that the strength of direct
$\gamma\pi\pi$ coupling
(let us call it provisionally $c$) is tighly connected with
universality violation (and then to a small violation of the
KSFR relation) by

\be
c=\displaystyle 1-\frac{g_{\rho \gamma} g_{\rho \pi \pi}}{m_{\rho}^2}
\label{resbando}
\ee

This relation  is found to give  $c=-0.182 \pm 0.008$ when
the constraint $c \equiv 1-a/2$ is forced. As a matter of
check, we have rerun our fit procedure by decoupling $c$
from $g_{\rho \gamma}$ and $g_{\rho \pi \pi}$. This corresponds
exactly to the model M$_3$ of \cite{benayoun2}, however leaving
free $g_{\rho \gamma}$ as it should. We have, of course,
obtained a very good fit ($\chi^2/{\rm dof}=66/76$), fixing also
$\lambda=1$, which provides $ c=-0.171 \pm 0.018$,
in fairly good agreement with the HLS model result{\footnote {
In \cite{benayoun2}, with $g_{\rho \gamma}$ fixed at its PDG
value and $\lambda$ left free,
the result was $c=-0.26 \pm 0.04$; with this respect,
note a misprint in the caption of Table 1 here: $A=c_0+c_2 m^2$
should be read   $A=-(c_0+c_2 m^2)$.}}.  In other
words, the HLS model is able to provide a meaning to
the  direct coupling $c$ in terms of only the $\rho^0$
parameters (see Eq.(\ref{resbando}) just above). This (non obvious)
correlation was, of course, completely missed in \cite{benayoun2},
and can be considered as a remarkable success of the HLS model.
Finally, the failure of VMD1 and WCCWZ tends also to support
the conclusion that $g_{\rho \gamma}$ is more consistent with
a constant value, than carrying a $q^2$ dependence. Indeed,
this $q^2$ dependence produces a too strong suppression of
the $\rho$ contribution when going down to threshold.

The question is now how to compare the other $\rho^0$ parameters
given above to the corresponding existing measurements \cite{PDG}.
The main problem
with an object as broad as the $\rho^0$ meson is that its shape
and, correspondingly, its observed parameters are highly influenced
by  phase space effects and by the production mechanism (creation
amplitude). This explains the wide spectrum of values reported in
the Review of Particle Physics \cite{PDG}.
Generally, phase space effects are perfectly known
and departures from expectations have a physical meaning, as in
the decay $\eta' \ra \pi^+ \pi^- \gamma$ (see Ref. \cite{benayoun2,cbar}
for instance). In hadronic  and in photoproduction experiments, one
has to rely on expressions for the production amplitudes
$\pi N \ra \rho N$ and  $\gamma N \ra \rho N$, which are actually
guesses to a large extent; Ref. \cite{PRoos} illustrates the dependence
of the $\rho$ mass on several (and all reasonable) guesses for the
$\pi N \ra \rho N$ amplitude. In the case of $e^+e^-$ annihilations,
we also work under assumptions which can be discussed (see Ref.
\cite{benayoun,benayoun2} for instance) mainly about the transition
$\gamma \ra \rho^0$ and the possible existence of a constant
$\gamma \pi^+ \pi^-$ coupling. One of the original motivations of
the present work was indeed to rely on both $e^+ e^- \ra \pi^+ \pi^-$
and  $e^+ e^- \ra \mu^+ \mu^-$ data in order to get rid of a
possible $\gamma \pi^+ \pi^-$ coupling. We have shown that the single
existing data set with $\mu \mu$ final state is  not precise enough
in the $\rho^0$ region
in order to achieve this program. However, it happens that the existing
information on $\delta_1^1$, obtained in a completely independent way,
 allows us to single out the influence of the
$\gamma \ra \rho^0$ transition and those of the $\gamma \pi^+ \pi^-$
coupling. The quality of the comparison between the phases of the
HLS  model on the one hand and the data points of Ref.
\cite{petersen} on the other hand
shows that these two sources of errors are well controlled.

The fit values obtained for our free parameters and given in
Eq.~(\ref{respdg}) are calibrated by a fit on \ep data, {\it i.e.}
they optimise
only $|F_{\pi}(q^2)|$. The expression deduced for $\arg{[F_{\pi}(q^2)]}$
automatically matches all known information on the phase $\delta^ 1_1$
without any further tuning,
as illustrated in Sections \ref{phaseshift} and \ref{threshold}.
Therefore, we conclude that our results in Eq.~(\ref{respdg})
are little affected by systematic errors due to modelling. The single
remaining freedom in defining the $\rho^0$ parameters is the definition
used for mass and width. What we have quoted corresponds to the
most usual Breit--Wigner definition.

Moreover, ChPT predictions \cite{knecht,gl85b}
near the two--pion threshold
allow us to check even more accurately the quality of the information
deduced from $e^+ e^-$ data.
We can therefore conclude that
the HLS model carries information reliable enough
to lessen to a large extent the dependence of the $\rho^0$
parameters on the production mechanism.

An improvement over the results
given in Eq.~(\ref{respdg}) would require comparison with other independent
data and an updating of the Froggatt--Petersen spectrum \cite{petersen}, using
more recent data and a more accurate $\rho$ pole position \cite{bernicha}.
Moreover, the accuracy of the existing data on \ep and \emuon
is limited by large systematic uncertainties. One can expect a dramatic
improvement from the new experiment with the {\sc cmd2} detector, now
in progress at Novosibirsk, aiming to reach a statistical accuracy
of 3\% per point and an overall systematic uncertainty better than 1\%
\cite{khazin}.

Finally, our results suggest
a HLS-like model might also prove useful in parametrising
hadronic spectral functions measured in $\tau$ decay where, to this
point, VMD2-like parametrisations have typically been used \cite{aleph}.


\section{Summary and Conclusions}
\label{conclusion}

We have studied a
 variety of vector meson dominance models
in both non--unitarised and unitarised forms. They depend on a few
parameters (mass of the $\rho$ meson, its coupling constants to
$\pi \pi$ and $e^+ e^-$, the shape parameter $\lambda$ and the Orsay phase
needed in order to describe the $\rho-\omega$ mixing). We have fitted these
to both $e^+e^-\rightarrow \pi^+\pi^-$ and $e^+e^-\rightarrow \mu^+\mu^-$.
In order to study the behaviour of each solution, we have studied
how they match the $\pi \pi$ phase shift obtained under general
model independent assumptions from threshold up to 1 GeV. We have
also examined the value they provide for threshold parameters
($F_{\pi}$ at threshold and the scattering length $a^1_1$),
which can be estimated accurately from ChPT.

This represents the largest set of independent data
and cross--checks done so far. It happens that,
of the models considered, only the
unitarised HLS model is able to account for all examined
effects. We also find that the standard value
$\lambda=1$, corresponding to a point-like $\rho\pi\pi$ coupling,
is well accepted by the data for $\rho$ parametrisation.

Unlike the standard formulation of VMD, fits with this model
return a significant
non--resonant
contribution to the electromagnetic pion form factor.
This was found to have a value
$\simeq -e/6$.  This term is governed
by a small universality violation which changes
the HLS parameter $a$ from 2 to 2.4. All other models considered,
even if { they are} able to describe
$e^+e^-$ annihilations quite well,
are unable to account satisfactorily for the other available information.
It should be mentioned that, within the class of models considered,
our results tend to favor a constant
$g_{\rho \gamma}$ over a $q^2$ dependent one.

We give the  values
for the $\rho^0$ mass and for its partial
widths to $\pi^+ \pi^-$ and $e^+e^-$, obtained using the HLS model
with $\lambda=1$. This corresponds to describing the resonance mass
spectrum through the usual Breit--Wigner expression ($\lambda=1$).
We also obtain an estimate of Br$(\omega \ra \pi^+ \pi^-)$.

As such, we conclude that, of the models considered, the HLS model with
$a \simeq 2.4$ is the most favoured version for
implementing the VMD ansatz. Thus, it is interesting to consider
whether the success of its predictions for the magnitude
and phase of $F_\pi(q^2)$, could be
obtained in a way which does not need the assumption that the $\rho$ is
a dynamical gauge boson of a hidden local gauge symmetry.
Hence, looking for other models able to describe,
as successfully as the HLS model, the same large set of data is
useful in order to know whether the conceptual
motivation for this model { should be interpreted as having any
underlying significance.}

\vspace{1cm}

\begin{center}
{\bf Acknowledgements}
\end{center}
We would like to thank J.-L.~Basdevant, S.~Gardner,  M.~Knecht,
J.~Petersen, J.~Stern and A.W.~Thomas for helpful
discussions. This work is supported by the Australian Research Council
and the US Department of Energy under grant DE--FG02--96ER40989.
It is also supported by Natural Sciences and Engineering
Council of Canada.

\end{document}